\documentclass{elsart}
 \usepackage[italian,english]{babel}
 \usepackage{textcomp}
 \usepackage{amssymb}
 \usepackage[T1]{fontenc}
 \usepackage[dvips]{graphicx,color}
 \usepackage{epic}
 \usepackage{eepic}

\usepackage{graphics}
\usepackage{graphicx}
\usepackage{epsfig}
\usepackage{amssymb}

\graphicspath{{eps_figs/}}

\def\A{\textsf{A}}

\begin{document}

\begin{frontmatter}

\title{First results from a Liquid Argon Time Projection Chamber in a Magnetic Field}

\def\A{\kern+.6ex\lower.42ex\hbox{$\scriptstyle \iota$}\kern-1.20ex a}
\def\E{\kern+.5ex\lower.42ex\hbox{$\scriptstyle \iota$}\kern-1.10ex e}

\author{A.~Badertscher}{,}
\author{M.~Laffranchi}{,}
\author{A.~Meregaglia}{,}
\author{A.~M\"uller}{,}
\author{A.~Rubbia}

\address{Institute for Particle Physics, ETH H\"onggerberg, CH-8093 Z\"urich, Switzerland}

\begin{abstract}

A small liquid argon Time Projection Chamber (LAr TPC) was operated for the first time in
a magnetic field of 0.55 Tesla. The imaging properties of the detector were not affected
by the magnetic field. In a test run with cosmic rays a sample of through going and
stopping muons was collected. The chamber with the readout electronics and the
experimental setup are described. A few selected events were reconstructed and analyzed
and the results are presented. \\
The magnetic bending of the charged particle tracks
allows the determination of the electric charge and the momentum, even for particles not
fully contained in the drift chamber. These features are e.g. required for future
neutrino detectors at a neutrino factory.
\end{abstract}

\end{frontmatter}

\section{Introduction}
\label{sec:Intro}
\subsection{The liquid argon Time Projection Chamber}
The liquid argon time projection chamber (LAr TPC) was
proposed by C.~Rubbia in 1977 \cite{Rubbia77}. With an extensive R\&D program, including
the construction of several prototypes of increasing mass, the ICARUS collaboration
\cite{intro4} has demonstrated the feasibility of this novel technology for large mass
detectors.  A 600 ton (T600) module consisting of two identical 300 ton half-modules was
built and successfully tested
\cite{icarus04}. \\
The LAr TPC is a 3-dimensional homogeneous tracking device with excellent, high
resolution imaging properties, and, at the same time it is a fine grain calorimeter for
fully contained particles due to the measurement of the energy loss $dE/dx$ along the
charged particle tracks. The detector is ideally suited for neutrino physics and a
sensitive search for nucleon decays \cite{Rubbia:2004yq}. In ultra-pure LAr
(contamination < 0.1 ppb O$_2$ equivalent) the ionization charge of a track can be
drifted undistorted over a distance of the order of meters in a uniform
electric field. \\
 \begin{figure}[h]
 \begin{center}
 \hspace{-0cm}
   \includegraphics[width=9cm]{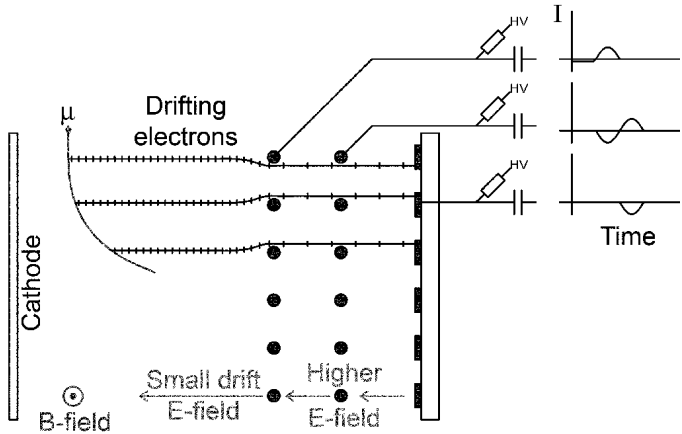}
 \caption[Schematic view of a TPC.]{\label{ionizchamber} Schematic view of a TPC in a magnetic
 field.  A charged particle is passing through  an ionization medium (e.g. LAr) and the electrons drift along
 the electric field lines to the sensors.}
 \end{center}
 \end{figure}
Figure~\ref{ionizchamber} shows a schematic view of a TPC; it consists of a cathode, the
sensor planes and field shaping electrodes (not shown in the figure) to produce a
homogeneous drift field. The ionization electrons drift along the electric drift field to
the sensor planes, where the charge is measured. In the LAr TPC described here there are
three sensor planes: the first two planes are multi-wire proportional chambers and the
third plane is a printed circuit board (PCB) with conducting strips. Preliminary results
obtained with this chamber were published in~\cite{Badertscher:2004py} and a detailed description is
given in \cite{dissmarco}. The sensor planes are biased with potentials, such, that the
wire chambers are transparent to the drifting electrons, they only pick up an induced
signal from the electrons passing through the plane (induction planes), the electrons are
then collected on the strips of the third plane (collection plane). Each plane provides a
two-dimensional projection of a track with the number of the hit wire (strip) and the
drift time as coordinates. The $t_0$ for the drift time is given by trigger counters. The
drift distance is calculated from the measured drift time and the known drift velocity
$v_d$. Combining the wire hits of equal drift time from at least two planes allows the
three--dimensional reconstruction of a track~\cite{icarus04,Rico02}.

As a consequence of the short mean free path of the drifting electrons in LAr, they do
not gain enough kinetic energy between collisions to ionize other atoms, thus, there is
no charge multiplication at the wires in LAr. However, the high ionization density in the
liquid allows a direct measurement of the ionization charge with a very low noise
amplifier; the expected charge for a minimum ionizing particle (mip) is of the order of
$13\,000$~electrons for $2$\,mm track length (at a drift field of $0.5$\,kV/cm). The
commercial VME--like CAEN modules\footnote{www.caen.it}, designed and built for the
ICARUS experiment (see chapter~\ref{readout}), have been used as front end electronics. \\
The low drift velocity ($1.6$\,mm/$\mu$s at a drift field of $0.5$\,kV/cm) and the long
drift paths in detectors with big sensitive volumes (as e.g. in ICARUS) implies that the
electrons spend a long time in liquid argon. This leads to the requirement of an
extremely low concentration of electronegative impurities (as e.g. oxygen) in LAr because
the drifting electrons can be captured by electronegative atoms or molecules, decreasing
the charge arriving at the sensors. If the impurity concentration is constant over the
whole volume, the charge decreases exponentially with the drift time:
 \begin{equation}
 Q(t)=Q(t_0)\:e^{-\frac{t}{\tau}}
 \label{puritydef}
 \end{equation}
where $\tau$ is the mean lifetime of the electrons in argon. The lifetime is directly
connected to the $\mathrm{O_2}$ equivalent impurity concentration $\rho$ by an inverse
linear relationship \cite{Buckley89}:
 \begin{equation}
 \tau[\mathrm{\mu s}] \approx \frac{300}{\rho [ppb]}
 \label{purityconvertion}
 \end{equation}
The commercial LAr 48 is specified to $1$\,ppm; this corresponds to a mean lifetime of
only $0.3$\,$\mu$s, absolutely insufficient for a TPC. A considerable amount of R\&D has
been performed by the ICARUS collaboration in order to
master the argon purification process \cite{icarus04,Bettini91}. \\
The drift velocity of the electrons in liquid argon is small compared to the thermal
velocity as a result of frequent collisions of the electrons with the argon atoms; it can
be approximated by \cite{Sauli77}
\begin{equation}
\label{drifteq1} v_d = \frac{e}{2m} \cdot E \cdot \tau_c,
\end{equation}
where $\tau_c$ is the mean time between collisions. The drift velocity has been measured
at different temperatures as a function of the drift field, and was fitted to an
empirical formula \cite{nima516}. For later use (see
chapter~\ref{lorentz-driftvel-diff}) we introduce here the mobility of the electrons,
defined as \cite{Sauli77}:
 \begin{equation}
 \mu \equiv \frac{v_d}{E}
 \label{defmobility}
 \end{equation}
Except for very low E-fields, the electron mobility is not constant. With
equation~\ref{drifteq1}, we find for the mobility:
 \begin{equation}
 \mu=\frac{e}{2m}\cdot \tau_c
 \label{mobility1}
 \end{equation}

\subsection{The LA\MakeLowercase{r} TPC in a magnetic field}
\label{lorentz-driftvel-diff} The possibility to complement the features of the LAr TPC
with those provided by a magnetic field has been considered and would open new
possibilities \cite{Badertscher:2004py,Rubbia:2001pk,Rubbia:2004tz,Bueno:2001jd}: (a) charge discrimination, (b)
momentum measurement of particles escaping the detector (e.g. high energy muons), (c)
very precise kinematics, since the measurement precision is limited by multiple
scattering (e.g. $\Delta p/p$ 4\% for a track length of $L\!=\!12$\,m and a field  of
$B\!=\!1$\,Tesla). The challenging possibility to magnetize a very large,
multi--kton, volume of argon has been addressed in \cite{cline03}.\\
The orientation of the magnetic field can be chosen such, that the bending direction is
in the direction of the drift where the best spatial resolution is achieved. This is
possible since the Lorentz angle is small, as is shown below. However, it is not
mandatory and the B--field could also be parallel to the drift field. In the following we
consider the case where the magnetic field is perpendicular to the electric drift
field. \\
The required magnetic field for charge discrimination for a thickness $x$ in LAr
($x_p$ is the projected thickness into the bending plane) is given by the bending
parameter $b$ \cite{Rubbia:2004tz}, as defined in Figure~\ref{def-bending}:
 \begin{figure}
 \begin{center}
    \includegraphics[width=9cm]{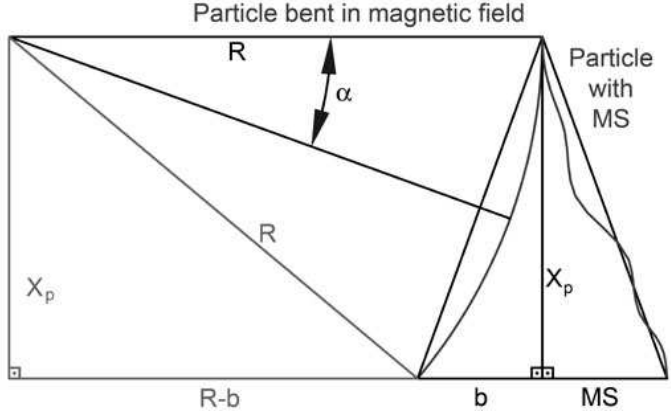}
  \caption[Definition of the bending parameter of a particle.]
 {\label{def-bending} Definition of the bending parameter $b$ of a particle,
  where $x_b$ is the thickness of the particle trajectory projected into the bending plane,
  and $R$ is the bending radius.}
 \end{center}
 \end{figure}
 \begin{equation}
 b^2-2Rb+x_p^2=0
 \label{bending1}
 \end{equation}
Neglecting the $b^2$ term for $\alpha\!\ll\!1$, $b$ can be approximated by:
 \begin{equation}
 b\approx\frac{x_p^2}{2R}
 \label{bending2}
 \end{equation}
The radius of curvature $R$ of the charged track due to the magnetic field is given by:
 \begin{equation}
 R=\frac{p_t}{(0.3\,B)}
 \label{curvature}
 \end{equation}
where $B$ is the magnetic field strength, $p_t$ is the transverse momentum, i.e.
$p_t=p\cdot\,\cos\!\lambda$, and $\lambda$ is the angle between the track and its
projection on the bending plane (pitch angle). Thus, the bending becomes:
 \begin{equation}
 b \approx \frac{x_p^2}{2R}=\frac{0.3\,B(Tesla)(x_p(m))^2}{2p_t(GeV)}
=\frac{0.3\,B(Tesla)(x(m))^2\,\cos\!\lambda}{2p(GeV)}
 \label{sagitta1}
 \end{equation}
The multiple scattering contribution to $b$ can be obtained from \cite[eq.
27.10]{PDBook}. Neglecting the logarithmic correction, inserting $X_0\!=\!14$\,cm as
radiation length of liquid argon and using the projected deviation $MS=x \theta_0
/\sqrt{3}$ (see also Figure~\ref{def-bending}), one obtains:
 \begin{equation}
 MS\approx \frac{0.02\:(x(m))^{3/2}}{p(GeV)}=\frac{0.02\:(x_p(m))^{3/2}}{p_t(GeV) \sqrt{\cos\!\lambda}}
 \label{MS}
 \end{equation}
The momentum resolution of a particle bending in a uniform magnetic field can be found
from equation~\ref{curvature}. With the introduction of the curvature, defined as
$k\!\equiv\!1/R$, it becomes:
 \begin{equation}
  \frac{\Delta p}{p}=\frac{p\,\Delta k \,\cos\!\lambda}{0.3\,B}
 \label{curvature1}
 \end{equation}
At low momenta, we can safely neglect the contribution from the position measurement
error given the readout pitch and drift time resolution and use the \cite[eq.
28.44]{PDBook} for the $\Delta k$ due to the multiple scattering. The momentum resolution
is then given by
\begin{equation}
 \frac{\Delta p}{p}\approx \frac{0.14}{B(Tesla)\sqrt{(x(m))}\,\cos\!\lambda}=
 \frac{0.14}{B(Tesla)\sqrt{(x_p(m))\,\cos\!\lambda}}
 \label{momentumresolution}
 \end{equation}
and the statistical significance for charge separation can be written as ($b^\pm$ are the
bending for positive and negative charges):
 \begin{equation}
 \sigma\approx \frac{b^+\!-\!b^-}{MS}\approx \frac{2b}{MS}\approx 15\,B(Tesla)\sqrt{x(m)\,\cos^3\!\lambda}
 \label{momentumresolution1}
 \end{equation}
For example, with a field of $0.55$\,T, the charge of tracks of $10$\,cm length can be
separated at $2.6\,\sigma$ with $\lambda\!=\!0$. The requirement for a $3\sigma$ charge
discrimination can be written as $b^+\!-\!b^-\!=\!2b\!>\!3$\,MS, which implies a field
strength of
 \begin{equation}
 B\geq \frac{0.2\,(Tesla)}{\sqrt{x(m)\,\cos^3\!\lambda}}
 \label{Magnetrequirement}
 \end{equation}
For long penetrating tracks like muons, a field of $0.1$\,T allows to discriminate the
charge for tracks longer than $4$\,m, if perpendicular to the magnetic field. This
corresponds for example to a muon momentum threshold of $800$\,MeV/c. \\
Unlike muons or hadrons, the early showering of electrons makes their charge
identification difficult. The track-length usable for charge discrimination is limited to
a few radiation lengths after which the showers make the recognition of the parent
electron more difficult. In practice, charge discrimination is possible for high fields
$x\!=\!1X_0\!\rightarrow\!B\!>\!0.5$\,Tesla, $x\!=\!2X_0\!\rightarrow\!B\!>\!0.4$\,Tesla,
$x\!=\!3X_0\!\rightarrow\!B\!>\!0.3$\,Tesla. From simulations, we found that the
determination of the charge of electrons of energy in the range between $1$ and $5$\,GeV
is feasible with good purity, provided the field has a strength in the range of $1$\,T.
Preliminary estimates show that these electrons exhibit an average curvature sufficient
to have electron charge discrimination with an efficiency of $20$\,\% for a contamination
with the wrong charge of less than $1$\,\% \cite{Bueno:2001jd}.

\begin{figure}
 \begin{center}
 \hspace{-0cm}
  \includegraphics[width=8cm]{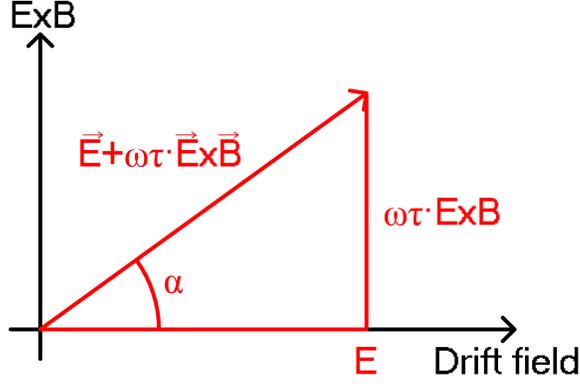}
  \caption{\label{Lorentz1} Definition of the Lorentz angle $\alpha$ (for perpendicular E-- and B--fields).}
 \end{center}
 \end{figure}

In the presence of a magnetic field a Lorentz force is acting on each moving charge,
modifying the drift properties of the electrons. As a consequence, the electrons will not
move along the electric field lines, but on a straight line at an angle $\alpha$ to the
E--field. In general, the drift velocity in a static electromagnetic field is given by
\cite[eq. 24.7]{PDBook}:
 \begin{equation}
 \overrightarrow{v}_{DB}=\frac{v_d}{1+\omega^2\tau_c^2}[\widehat{E}+\omega\tau_c(\widehat{E}\wedge\widehat{B})+
 \omega^2\tau_c^2(\widehat{E}\cdot \widehat{B}) \widehat{B}]
 \label{vdriftmag}
 \end{equation}
where $\widehat{E}$ is the unit vector in the E--field direction, $\widehat{B}$ the unit
vector in the B--field direction and $\omega$ is the cyclotron frequency:
 \begin{equation}
 \omega=\frac{e \cdot B}{m}
 \label{cyclotron}
 \end{equation}
In the case that the electric and magnetic fields are perpendicular to each other, the
electrons will drift along a straight line at an angle $\alpha$ with respect to the
electric field lines (see Figure~\ref{Lorentz1}). In this case equation~\ref{vdriftmag}
is simplified to:
 \begin{equation}
 \overrightarrow{v}_{DB}=\frac{v_d}{1+\omega^2\tau_c^2}[\widehat{E}+\omega\tau_c(\widehat{E}\wedge\widehat{B})]
 \label{vdriftmagperp}
 \end{equation}
From Figure~\ref{Lorentz1} it can be seen that for the case of perpendicular E-- and
B--fields the Lorentz angle $\alpha$ is given by:
 \begin{equation}
 \tan \alpha=\omega\tau_c
 \label{Lorentz2}
 \end{equation}
From equation~\ref{mobility1} we obtain for $\tau_c$:
 \begin{equation}
 \tau_c=\mu \cdot \frac{2m}{e}
 \label{mobility2}
 \end{equation}
With the definition of the mobility in equation~\ref{mobility1} and using the empirical
expression for the drift velocity \cite{nima516}, we can calculate the Lorentz angle for
a given magnetic field (for perpendicular E-- and B--fields):
 \begin{equation}
 \tan\!\alpha=\frac{e\cdot B}{m}\cdot\mu\cdot\frac{2m}{e}=2\mu B
 \label{Lorentz3}
 \end{equation}
The Lorentz angle is expected to be very small in the liquid, e.g. $1.7^{\circ}$ at $E =
500$\,V/cm$^{-1}$ and $B=0.5$\,T. Embedding the volume of argon into a magnetic field
should therefore not alter the imaging properties of the detector and the measurement of
the bending of charged hadrons or penetrating muons would allow a precise determination
of the momentum and a determination of their charge.

\section{TPC construction}
\label{sec:construction}
\subsection{Overview}
 \begin{figure}
 \begin{center}
    \includegraphics[width=12cm]{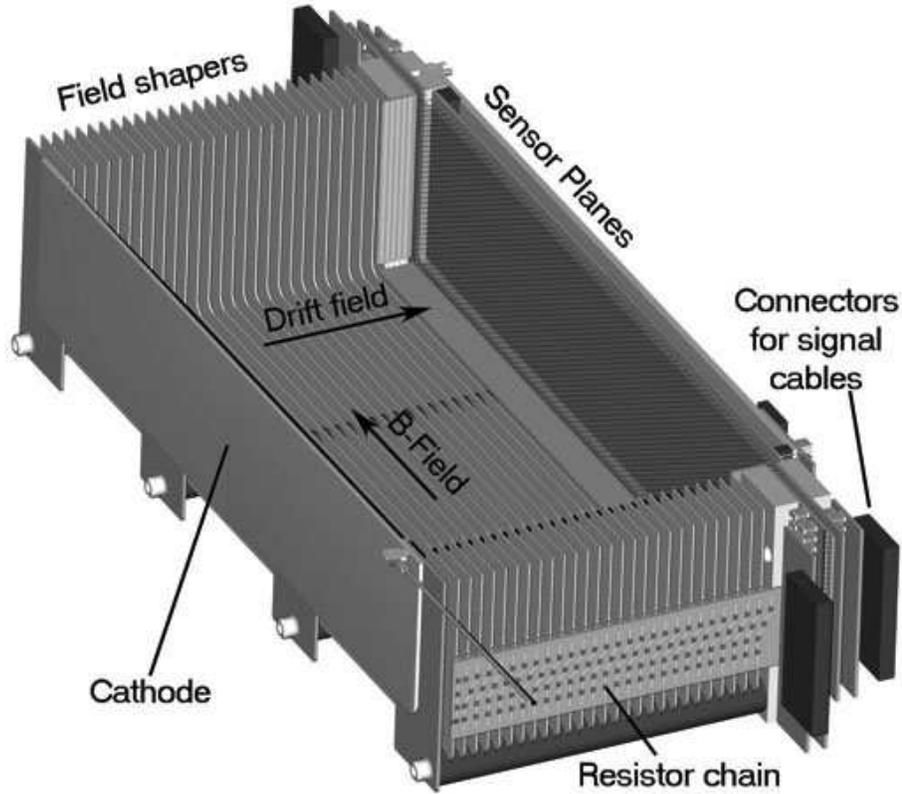}
 \caption{\label{chambercut} Cut view into the open drift chamber.}
 \end{center}
 \end{figure}
 \begin{figure}
 \begin{center}
    \includegraphics[width=12cm]{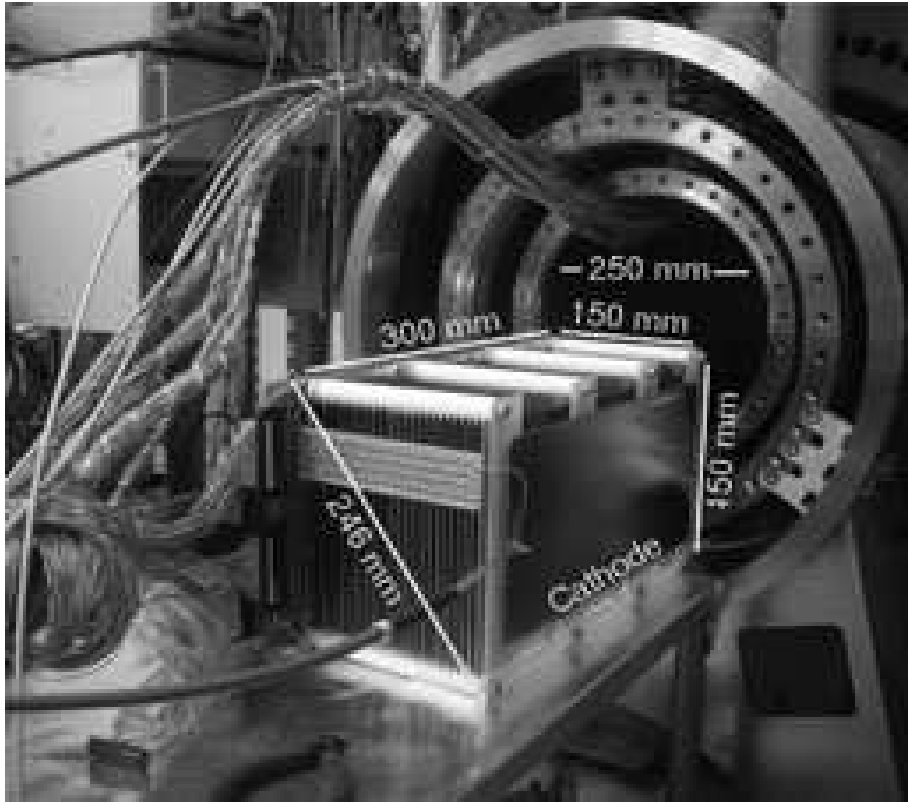}
  \caption[Picture of the liquid argon time projection chamber.]
 {\label{photochamber} The liquid argon time projection chamber ready to slide into the cryostat.}
 \end{center}
 \end{figure}
The dimensions of the cryostat and the chamber were chosen to fit into the recycled
SINDRUM I magnet\footnote{The magnet was kindly lent to us by the Paul Scherrer Institute
(PSI), CH-5232 Villigen, Switzerland.}. The sensitive volume of the chamber has a length
(in the direction of the solenoid axis) of $300$\,mm, a height of $150$\,mm and a width
corresponding to a maximal drift length of
$150$\,mm.  \\
The drift chamber (see Figure~\ref{chambercut}) consists of a rectangular cathode, $27$
field shaping electrodes equally distributed over the whole drift path and three detector
planes. The first two detector planes are wire chambers with the wires oriented at
$\pm60^\circ$ to the vertical; the $127$ stainless steel wires for each plane have a
diameter of $100$\,$\mu$m and a pitch of $2$\,mm. The third plane is a PCB with
horizontal strips with a width of $1$\,mm and a pitch of $2$\,mm on which the drift
electrons are collected. The potentials of the three planes can be varied from outside;
the wires and the strips are coupled through a $470$\,M$\Omega$ resistor to the bias high
voltage and a $1.4$\,pF capacitor to $3$\,m long twisted pair cables (see
chapter~\ref{readout}), which are connected to the feedthroughs at the warm part
of the cryostat.\\
The cathode and the field shaping electrodes are designed to hold the high voltage (up to
a maximum of $22.5$\,kV was applied) and are thus carefully insulated with PCBs and
Macor\texttrademark$\:$rods. Eight Macor rods with a diameter of $15$\,mm at the bottom
and the top of the chamber were used to assemble the cathode, the field shaping
electrodes and the the sensor planes (see Figure~\ref{photochamber}), giving enough
mechanical stability to the chamber. With a resistor chain, connected to the cathode HV
at one end and to the wire chambers at the other end, the field shaping electrodes are
connected to the appropriate potential, in order to obtain a homogeneous drift field in
the whole chamber volume. Figure~\ref{photochamber} shows the assembled chamber ready to
slide into the cryostat.

The chamber is oriented in the LAr vessel such, that the drift direction is horizontal;
the $\vec{E}$--field is perpendicular to the $\vec{B}$--field, which is also horizontal.
This orientation of the chamber has the advantage that the bending of the charged
particles incident from the top is in the drift
direction, where the spatial resolution is best. \\
Table~\ref{tablesummary} summarizes the most important chamber parameters.

 \begin{table}
 \begin{center}
 \begin{tabular}{|r|l|}
 \hline
 Total volume & $400\,mm \times 170\,mm \times 170\,mm$ \\ \hline
 Sensitive volume & $300\,mm \times 150\,mm \times 152.9\,mm$ \\ \hline
 Sensing planes & $1^{\mathrm{st}}$ Induction \\ \hline
                     & $2^{\mathrm{nd}}$ Induction \\ \hline
                     & Collection \\ \hline
 Induction planes & wires at $\pm\,60^\circ$ to the vertical \\ \hline
                     & $127$ wires per plane \\ \hline
                     & $2\,mm$ pitch \\ \hline
                     & $-200\,V$ bias voltage of $1^{\mathrm{st}}$ induction plane \\ \hline
                     & $0\,V$ bias voltage of $2^{\mathrm{nd}}$ induction plane \\ \hline
                     & $2.8\,mm$ distance between $1^{\mathrm{st}}$ and $2^{\mathrm{nd}}$ induction plane\\ \hline
 Collection plane & $1$\,mm wide copper strips on a printed circuit board \\ \hline
                     & $75$ strips (channels) \\ \hline
                     & $2\,mm$ pitch \\ \hline
                     & $+280\,V$ bias voltage \\ \hline
                     & $3.1\,mm$ distance between $2^{\mathrm{nd}}$ induction plane and strips \\ \hline
 Electronic channels & $127+127+75=329$ \\ \hline
 Electric drift field& $0<E<2\,kV/cm$ \\ \hline
 Expected charge & \\
  for a mip ($\mu$) & $\sim 13\,000$ electrons for $2$\,mm track length at $0.5\,kV/cm$\\ \hline
 \end{tabular}
 \caption{\label{tablesummary} Summary of the chamber parameters}
 \end{center}
 \end{table}

\subsection{The sensing planes}
\label{sensingplanes}

 \begin{figure}
 \begin{center}
    \includegraphics[width=16cm]{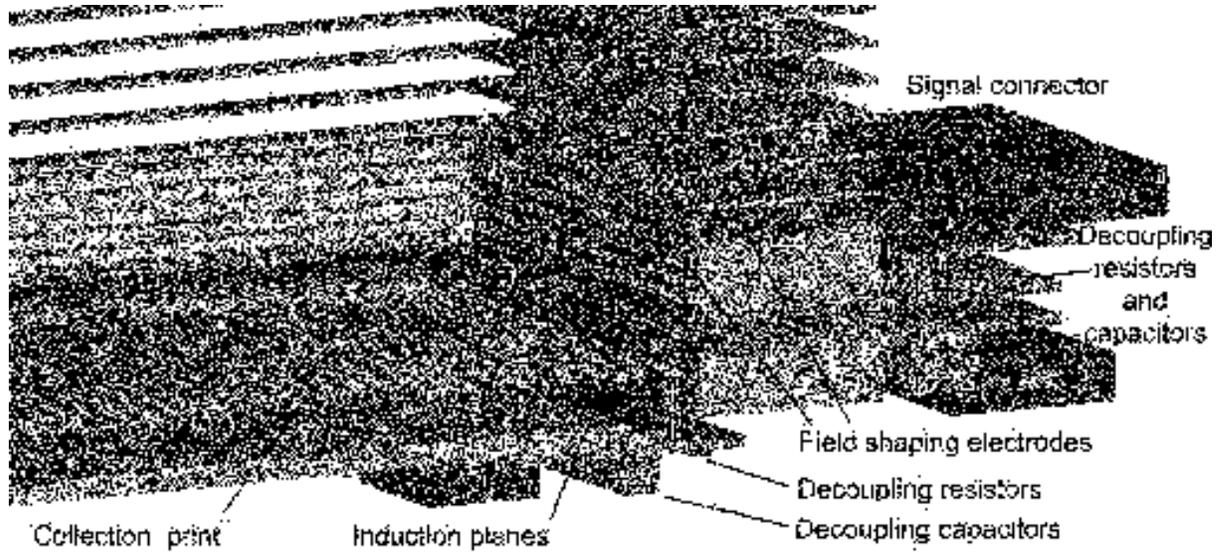}
  \caption[Zoomed view of the three sensing planes.]
 {\label{chambercutzoom} Zoomed view of the three sensing planes with the signal
 decoupling prints.}
 \end{center}
 \end{figure}
The induction planes are composed of a robust frame made of non-magnetic stainless steel
(316L) with inner dimensions of $302$\,mm\,$\times\,152$\,mm (outer dimensions:
$360$\,mm\,$\times\,194$\,mm). The frame thickness of the first induction plane is
$10$\,mm and for for the second induction plane it is $8$\,mm; it was calculated to avoid
a deformation of the frames under the force of the wires. Each plane has $127$ wires and
each wire has a tension (at room temperature) of ($0.75\,\pm\,0.03$)\,N; the distance
between the two wire planes was set to $2.8$\,mm with a plastic spacer. The PCBs to
solder the wires, the coupling capacitors and the resistors were glued to the steel frame
with Araldite\texttrademark. The $100$\,$\mu$m wires are made of the same material as the
frames (steel 316L); they were strung and soldered at PSI. \\
The coupling capacitors and resistors are mounted on the same PCB with the wires (see
Figure~\ref{chambercutzoom}). The connectors for the signal cables are on a separate PCB,
connected through the coupling capacitors. \\
The collection plane is a PCB with a size of $300$\,mm\,$\times$\,$150$\,mm in order to
fit inside the frame of the second induction plane. On the PCB there are $75$~strips,
$1$\,mm wide and with a pitch of $2$\,mm. On the back side of the PCB a $2$\,mm thick
steel plate is mounted at a distance of $5$\,mm from the PCB. The steel plate and the PCB
are glued to the Macor bars. The steel plate is connected to ground; it stabilizes
mechanically the chamber and shields the collection plane. The PCB with the coupling
resistors, capacitors and the signal connectors is glued to the bottom side of the steel
plate; the strips are connected with copper wires to the PCB. Figure~\ref{chambercutzoom}
shows a zoomed view of the chamber with the assembled sensing planes and the position of
the coupling resistors and capacitors. \\
Tests were made to prove that the glued PCBs do not detach from the frames and the
capacitors do not break due to the mechanical stress during the cooling down: the chamber
(without cathode and field shaping electrodes) was cooled down as quickly as possible,
but avoiding excessive thermal stress. For a cooling down time of about $10$~hours, the
maximal temperature difference between the frame and the surrounding temperature was
about $4$\,$^\circ$C. This value for the maximal temperature difference was kept as a
limit during the cooling down of the whole chamber in the cryostat.

\subsection{The cathode and the field shaping electrodes}
 \begin{figure}
 \begin{center}
 \setlength{\unitlength}{1mm}
 \begin{picture}(130,75)
    \put(0,0){\includegraphics[width=60\unitlength]{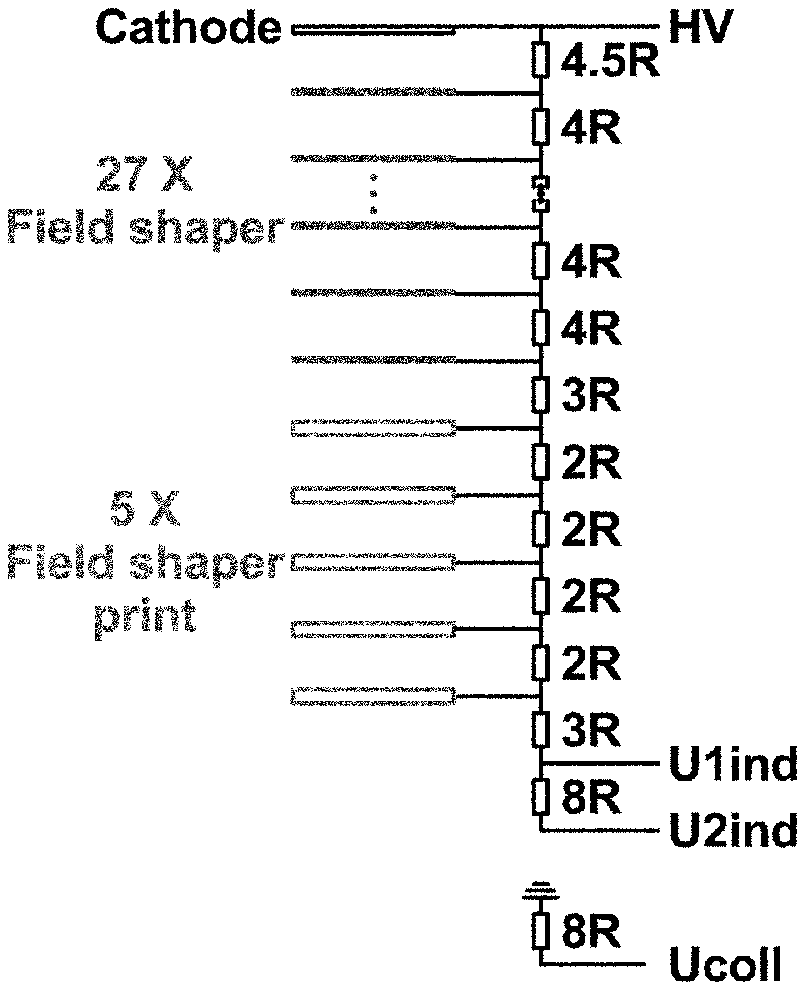}}
   \put(60,10){\includegraphics[width=60\unitlength]{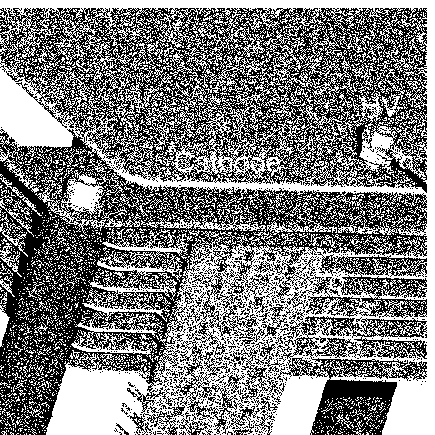}}
  \end{picture}
 \caption[Scheme and drawing of the resistor chain for the field shaping electrodes.]
 {\label{fieldshaperresistors} (left) Scheme of the resistor chain for the field shaping electrodes.
 (right) CAD drawing of the resistor chain connected to the field shaping electrodes.}
 \end{center}
 \end{figure}
The cathode is a $2$\,mm thick stainless steel plate connected to the high voltage and to
the resistor chain for the field shaping electrodes. The plate is big enough not to
distort the field in the drift volume near the cathode; it is insulated by a minimal
distance of $20$\,mm to the LAr vessel, which is grounded. This distance is enough for
liquid argon, which has a breakdown field strength of about
$1.1\!-\!1.4$\,MV/cm, depending on the purity of the argon. \\
The field shaping electrodes are made of $27$ stainless steel (316L) frames, $0.8$\,mm
thick, and surround the chamber volume. They are held at a distance of $5$\,mm by slits
in the Macor\texttrademark$\:$rods. The field shaping electrodes are connected to a
resistor chain from the high voltage to the wire chambers. The resistors are surface
mounted metal film resistors with a nominal value (at room temperature) of
$1$\,G$\Omega$. Figure~\ref{fieldshaperresistors}~(left) shows the electric scheme of the
resistor chain; the resistance between two field shaping electrodes is called 4R
(R\,=\,$1$\,G$\Omega$). The distance between the first induction plane and the closest
field shaping electrode is $16$\,mm. To avoid a distortion of the drift field near the
edge of this first wire plane, a PCB with strips (see Figure~\ref{chambercutzoom}) was
glued to the inner sides of the $10$\,mm thick wire frame and the strips were also
connected to the resistor chain; the resistance between the the first field shaping
electrode and the first strip is 3R, and between strips it is 2R.
Figure~\ref{fieldshaperresistors}~(right) shows how the resistor chain is connected
through gold plated pins to the field shaping electrodes. The total resistance of the
resistor chain is more than $100$\,G$\Omega$ and the current is negligible. The resistors
used for the chain were tested to withstand shock cooling in LN$_2$, since they were all
connected in series, hence, if one would break, there would be no drift field.

\subsection{The high voltage system}
The sensing planes are put at such a potential that the wire chambers are transparent for
the drifting electrons and then collected on the collection plane. With a wire diameter
of $100\,\mu$m and a pitch of $2$\,mm, the fields before ($E_1$) and after ($E_2$) the
plane have to fulfill the relation $E_2\!>\!1.37\!\cdot\!E_1$ in order that the wire
plane is $100$\,\% transparent \cite{dissmarco}. \\
During the test of the chamber different bias voltages were tested to increase the signal
to noise ratio; the planes were finally put to $-200$\,V for the first induction, $0$\,V
for the second induction and $+280$\,V for the collection plane, with a drift field of
$300$\,V/cm. This configuration corresponds to a field ratio of $2.3$ through the first
induction, and a ratio of $1.40$ through the second induction plane. \\
The potentials are set externally by a VME HV power supply\footnote{ISEG type VHQ 205L $2
\times 5$\,kV / $1$\,mA} with a declared ripple of less than $2$\,mV$_{pp}$. The
potentials were fed into the cryostat through $2\!\times\!20$\,kV SHV feedthroughs on a
CF40 flange. The second induction plane has a $5$\,kV SHV connector and it is closed with
a shortcut connector to the ground. The
feedthroughs are connected to the chamber by $3$\,m long coaxial cables (type RG58). \\
The cathode is supplied by a high voltage power supply\footnote{AIP WILD AG, type HCN
35--35000, $35$\,kV/$1$\,mA} with a declared ripple of less than $1\!\cdot\!10^{-5}$ or
$10$\,mV$_{pp}$. Since the argon gas has a low dielectric strength, the connection of the
HV feedthrough to the HV cable inside the argon container has been poured with
Araldite\texttrademark$\:$. During the test with cosmic ray muons the voltage of the
cathode was first set to $22.5$\,kV. After one week of operation breakthroughs occurred.
From that point on, the maximal voltage held was only $4.7$\,kV, corresponding to a drift
field of $300$\,V/cm.

\section{The readout electronics and the data acquisition system}
\label{readout}
 \begin{figure}[h]
 \begin{center}
    \includegraphics[width=15cm]{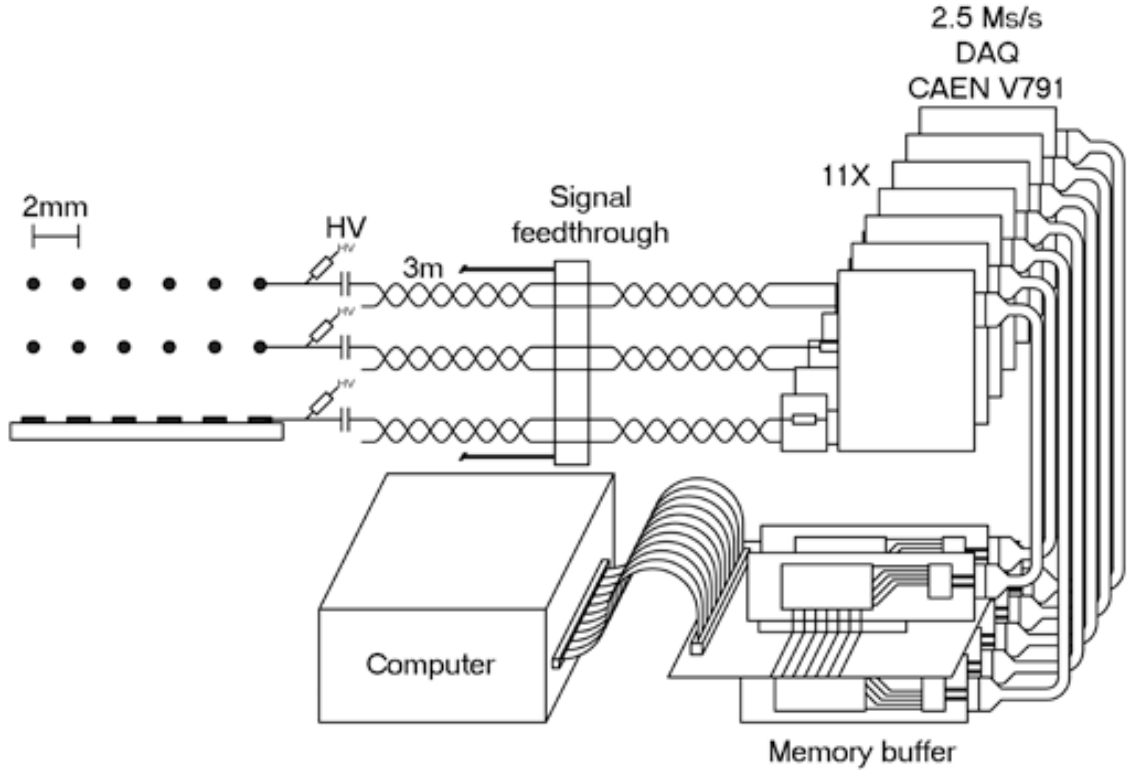}
  \caption{\label{schemeDAQ} Block diagram of the data acquisition system.}
 \end{center}
 \end{figure}
\subsection{The front--end electronics}
\label{frontend} The readout electronics works as a multichannel wave form digitizer for
the $329$~channels from the two induction planes and the collection plane. The V791
modules \cite{frontend} developed for the ICARUS experiment together with CAEN were used
as front--end electronics. Each channel is equipped with a low--noise current integrating
preamplifier \cite{preamp} followed by a Flash ADC with $10$~bit resolution. All channels
are sampled with a rate of $2.5$\,MHz, i.e., every $400$\,ns. The digitized data are
continuously stored in a circular buffer, which is large enough to contain the data of
all channels for a time interval of about $1$\,ms; the maximal drift time ($t_{dmax}$)
occuring during this experiment was only about $150$\,$\mu$s. Figure~\ref{schemeDAQ}
shows a schematic view of the data acquisition system. When a trigger occurs, the filling
of the buffer continues for at least $t_{dmax}$, in order to have all the samples of the
event stored in the buffer. Before the next trigger is accepted, all the data in the
buffer are transferred to a PCI computer card in a PC; the PCI card is read out with a
LabView\texttrademark$\:$program and the data stored on the
hard disk. \\
The signal with a charge corresponding to about $13\,000$~electrons is decoupled from the
high voltage by a surface mounted $470$\,M$\Omega$ resistor and a ceramic capacitor of
$1.2$\,nF (large compared to the input wire plus cable capacity of about $130$\,pF). The
chamber and the vacuum feedthroughs are connected by $3$\,m long twisted pair flat
cables. The twisted pair cables are Teflon insulated\footnote{Amphenol, type
425-3016-068, CERN catalogue 04.21.21.368.5} and are suited to use in vacuum and in high
purity liquid argon. The low capacity twisted pair cable also suppresses possible
microphonic noise due to vibrations in the magnetic field.  \\
Outside the cryostat about $20$\,cm of the same type of twisted pair cables connect the
feedthroughs to the front--end electronics. The signal cables are connected to the back
plane of a VME crate containing the CAEN V791 modules with the preamplifiers and the
ADCs. One module has 32 input channels. Its function is to amplify, shape and digitize
the signals coming from the detector and transmit the digitized data via a fast serial
link to the buffer module. Two different versions of the V791 module are available, the
Mod.~V791C and the Mod.~V791Q, which operate in "quasi current" and "quasi charge" mode,
respectively; the two versions differ in the feed back and restore time constants.
 \begin{figure}
 \begin{center}
    \includegraphics[width=12cm]{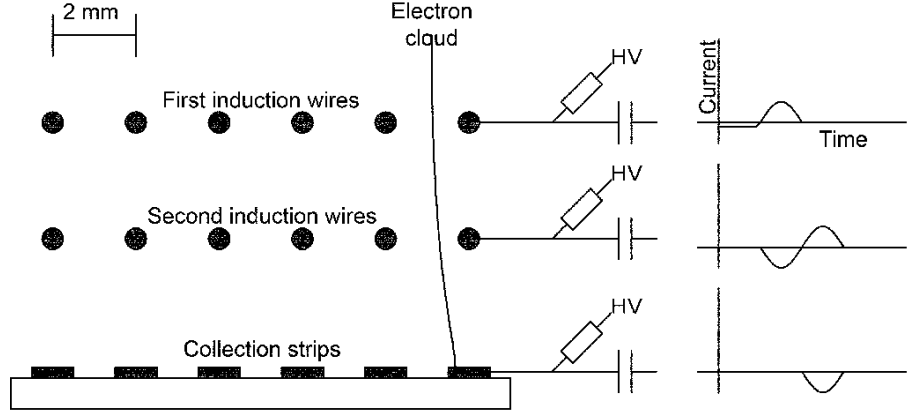}
  \caption{\label{pulses} Schematic view of the pulse shapes for the three sensor planes.}
 \end{center}
 \end{figure}
Figure~\ref{pulses} shows schematically the pulse shapes before the amplifier/shaper
stage for the three sensor planes; the current mode is used for the unipolar signals from
the first induction and the collection plane, and the charge mode is used for the middle
plane (second induction plane). The measured signals are induced by the moving charge of
the drift electrons and are proportional to the drift velocity. Thus, strictly speaking
the signal from the first induction plane is also bipolar: a small negative part from the
slowly drifting electrons approaching the plane from the drift volume, and a much larger
positive part induced when the electrons have passed through the plane and drift away
much faster in the larger drift field between the two induction planes. The bipolar
signal of the middle plane is caused by the electron cloud, which is first approaching
and then moving away from the plane. The collection plane, of course, produces a true
unipolar signal. \\
The digitized data are converted\footnote{National Semiconductor type DS90CR213} into
three LDVS (Low Voltage Differential Signaling) data streams and continuously sent to a
specially built interface with a circular buffer. At a clock frequency of 40 MHz, the
following $21$~bits of TTL data are transmitted with every clock cycle:
\begin{itemize}
\item 10--bit ADC data of the channels $k$;
\item 10--bit ADC data of the channels $k\!+\!16$;
\item 1 SYNC--bit, set to $1$1 when the ADC data of channels $15$ ($\mathrm{k}\!=\!15$)
and 31 are transmitted and $0$ otherwise.
\end{itemize}

 \begin{figure}
 \begin{center}
  \includegraphics[width=12cm]{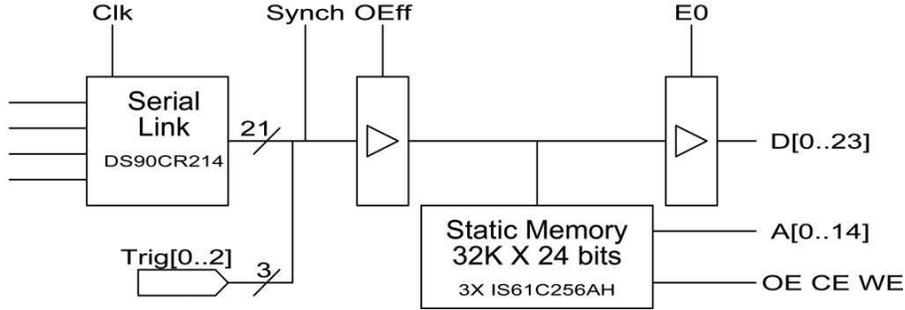}
 \caption{\label{memorybuffer} Schematic of the memory buffer card.}
 \end{center}
 \end{figure}

\subsection{The buffer module}
An interface between the eleven V791 CAEN cards and the computer has been built, able to
store the continuous flow of digital data in a circular buffer and to send the data to
the PCI card in the computer, once a trigger has occurred \cite{dissmarco}. The interface
consists of eleven buffer cards (one per V791 module) connected through a
dedicated bus to the main board.\\
Figure~\ref{memorybuffer} shows of the block diagram of the buffer card. A single buffer
card contains a serial link receiver circuit (DS90CR214 of National Semiconductor) to
convert the LVDS data streams from a V791 module back to $21$~bits of CMOS/TTL data. As
long as no trigger is detected, the data are stored into the memory. The buffer on each
card has a size of $32\,\mathrm{k}\,\times\,24$~bits. The two $10$--bit data samples plus
the synchronization bit transmitted in a cycle can hence be stored in a single memory
cell; there are $3$~bits left to store additional information, e.g. about the trigger.
The size of the memory is large enough to store $2048$ samples of a single channel. The
address of the memory is generated by the main board, using an increment on every clock
cycle and is the same for all the buffer cards. The address is coded with the Gray method
\cite{gray} in order to reduce the signal interference on the bus. \\
The main board consists of a bus with slots for $19$~buffer cards, two ALTERA FPGA (Field
Programmable Gate Array) chips and an independent static memory of $2\!\times 512$\,k$
\times 16$\,bit\footnote{ISSI, type IS61LV51216}. One ALTERA chip (the EPM3128ATC100-5)
is programmed to drive all the buffer cards during the waiting for a trigger and also
handles the trigger \cite{dissmarco}. After a trigger is detected, the EPM3128ATC100-5
chip waits a certain number of clock counts, which can be selected by a $5$ dip--switch,
and then stops the memory filling of all buffer cards in order to read the whole memory
and send it to the computer. Choosing with the switch an appropriate waiting time to stop
the memory filling after a trigger allows to set the length of the pre--trigger data,
which are stored with an event. This is important for the fit of the base--line of a
channel in the time before the actual signal starts. In our experiment with cosmic rays
the switch was set so that the trigger time corresponded to the memory address $1024$,
i.e. the middle of the buffer. Thus, half of the data stored as an event are pre--trigger
data.
 \begin{figure}
 \begin{center}
   \includegraphics[width=12cm]{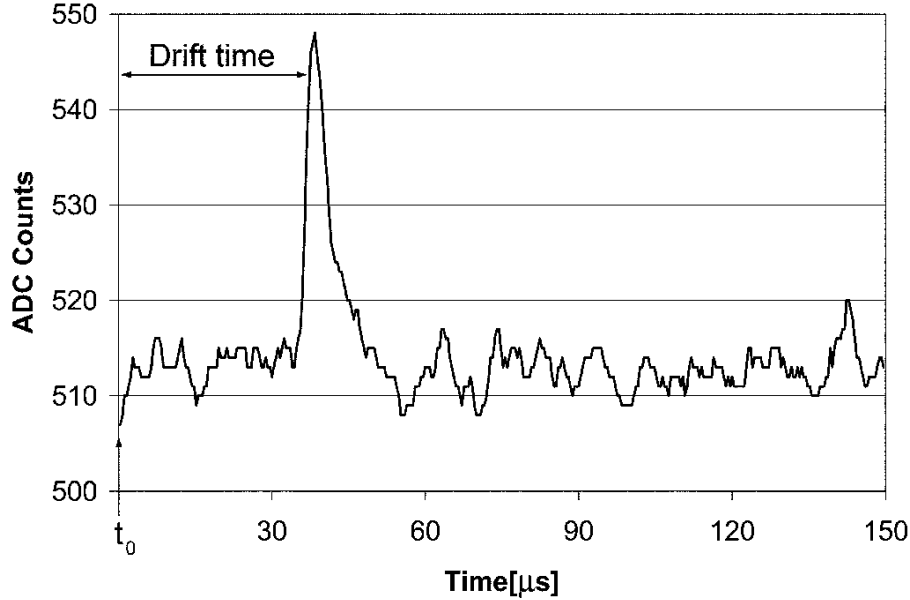}
  \caption[Raw data signal from the collection plane.]
 {\label{typicalsignal} Raw data signal from the collection plane; $t_0$ indicates the trigger time.}
 \end{center}
 \end{figure}

\begin{figure}
\begin{center}
 \includegraphics[width=10cm]{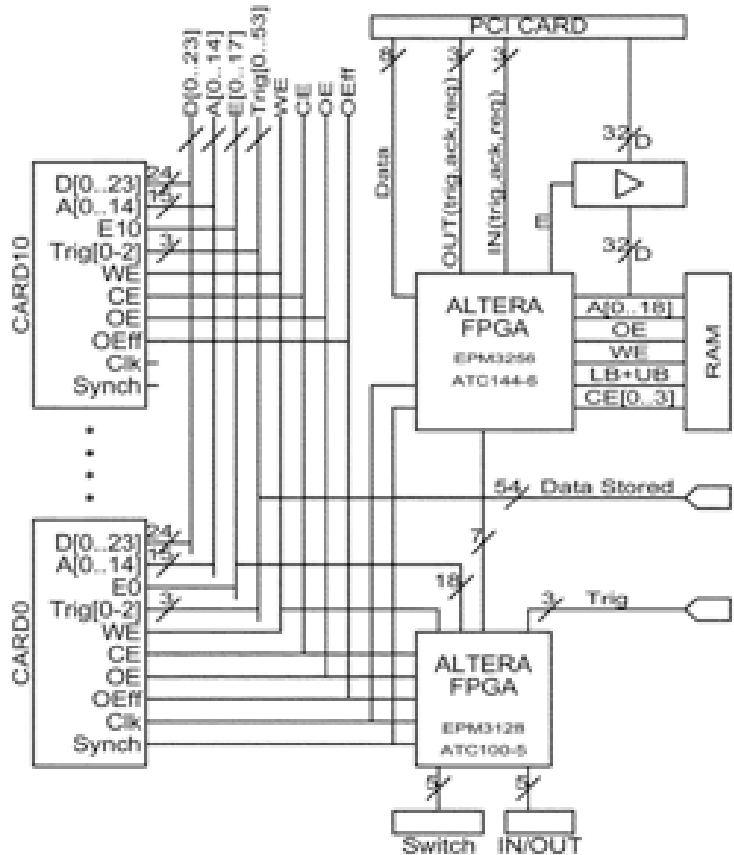}
\caption[Schematic of the interface module.]
 {\label{motherboard} Schematic of the buffer module.}
 \end{center}
 \end{figure}
Figure~\ref{typicalsignal} shows a stored signal from the collection
plane with the trigger position and the drift time indicated. \\
When the memory of the buffer card $0$ is completely read, the memory of the next buffer
card is read. The data are transmitted to the second ALTERA chip, the EPM3256ATC144-7,
and then to the computer via a PCI card. The second ALTERA EPM3256ATC144-7 is connected
with a $16\times 1$\,Mbits memory and could be programmed to make a first data
compression. In our experiment with the cosmic rays the ALTERA chip was only programmed
to transmit the data from the buffer card to the PCI computer card. \\
A LabView\texttrademark$\:$program running under Windows XP on a PC read the data from
the $12$\,MB/s PCI-7200\texttrademark$\:$from ADLink\footnote{ADLink Technology Inc.} and
stored them in a file on the hard disk; $100$~events were stored in a $137$\,MB file. The
LabView program also has the option to display online the signals of each event, but this
feature was normally deactivated during the data taking with cosmic rays, reducing the
dead time for the next trigger. The maximal trigger rate was in this case about
$1/2$\,Hz, corresponding to a data flow of about $685$\,kB/s.

\section{The experimental setup}
\label{setup}
 \begin{figure}
 \begin{center}
    \includegraphics[width=16cm]{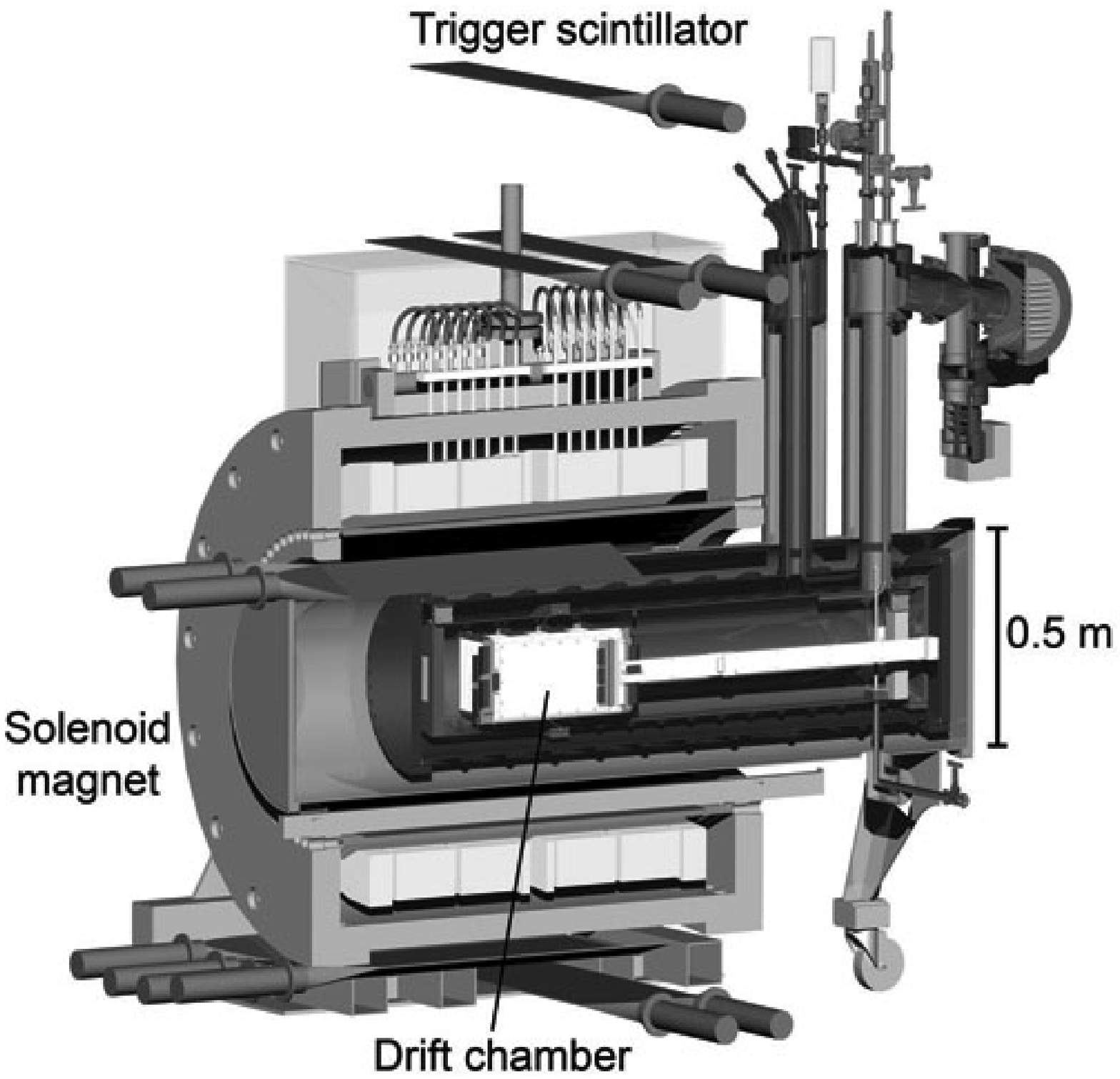}
  \caption{\label{allcutmu} Global view of the experiment.}
 \end{center}
 \end{figure}

 \begin{figure}
 \begin{center}
 \setlength{\unitlength}{1mm}
 \begin{picture}(134,95)
    \put(0,0){\includegraphics[width=79\unitlength]{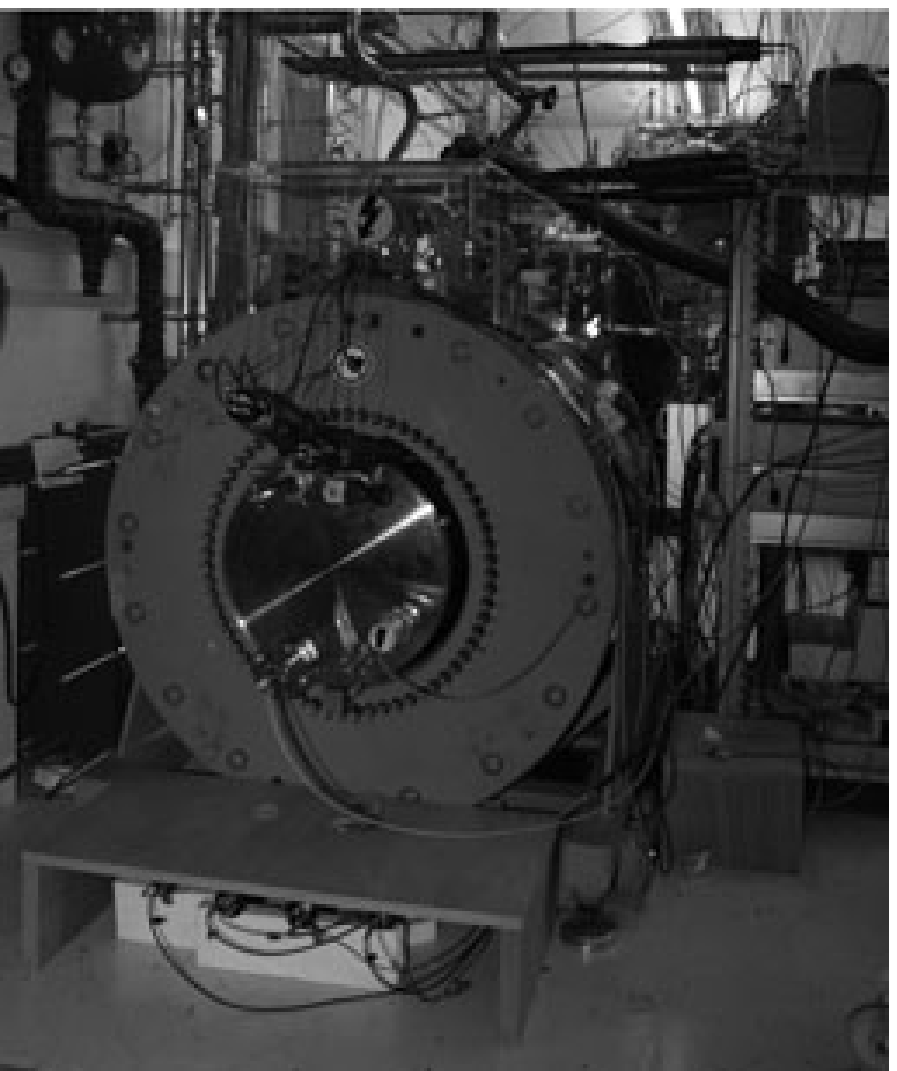}}
   \put(81,0){\includegraphics[width=53\unitlength]{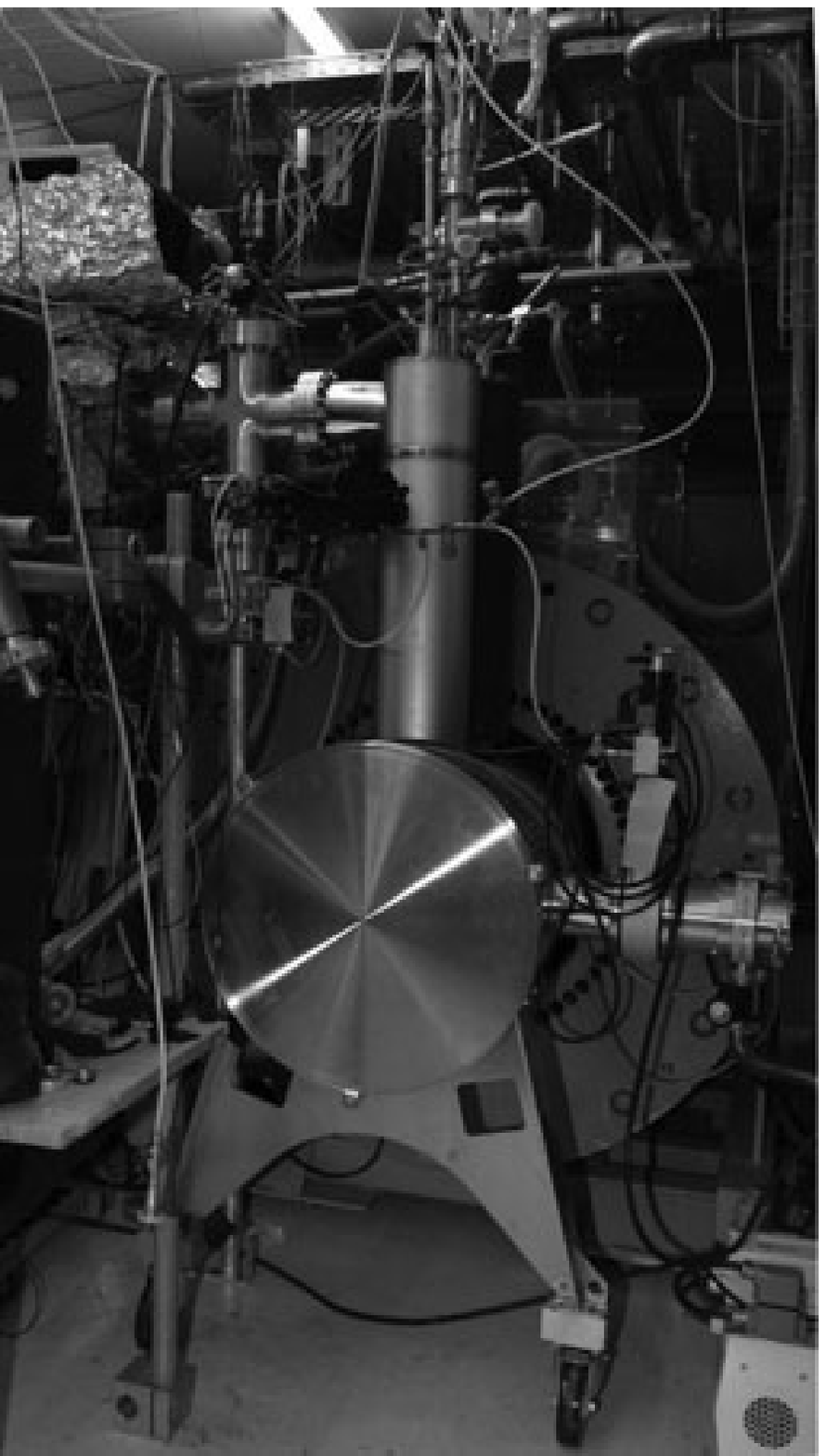}}
  \end{picture}
 \caption{\label{Photoexperiment} Pictures of the experiment taken from opposite sides.}
 \end{center}
 \end{figure}
The dimensions of the chamber and the LAr cryostat are such, that they fit into the bore
hole of the recycled SINDRUM I magnet from PSI. Figure~\ref{allcutmu} is a 3D CAD drawing
showing a cut through the setup with the essential components of the experiment. \\
The cryostat consists of three cylindrical stainless steel vessels (see
section~\ref{sectioncryostat}). The LAr vessel is shown in red and contains the chamber
and the purified liquid argon. The LAr is kept in a liquid nitrogen (LN$_2$) bath to
stabilize its temperature and pressure. The nitrogen bath, shown in blue, contains
nitrogen with an absolute pressure of about $2.7$\,bar in order not to freeze out the
argon, which is at about 1 bar. The outer most container for the vacuum insulation of the
cryostat is shown in green in Figure~\ref{allcutmu}. \\
The solenoid magnet can generate an axial magnetic field up to $0.55$\,T in the region of
the chamber (see section~\ref{sectionmagnet}). The magnet is cooled with a closed water
cooling system connected to the main laboratory water cooling circuit.\\
Eleven plastic scintillator counters are mounted on top, in the bore hole and underneath
the magnet as trigger counters for cosmic rays (see section~\ref{triggercounters}).\\
The temperature, the pressure and the level of both cryogenic liquids are monitored by
the slow control sensors and the data are stored in a file (see section~\ref{secslowcontrol}).\\
The commercial grade liquid argon is filled into the LAr container through a purification
cartridge (see Figure~\ref{Fillingfig}) in order to reach a contamination of the LAr with
electro-negative impurities of about $2$~ppb ($\mathrm{O_2}$ equivalent), which is
necessary to obtain a lifetime of the drift electrons of the order of the maximal drift
time, i.e., about $150$\,$\mu$s (see section~\ref{secpurification}). The LAr
container is pumped before it is filled with LAr. \\
The experiment is located $11$\,m underground in a laboratory of the Institute for
Particle Physics, ETH Zurich. The resulting underground depth of the chamber is $7.3\,m$
water equivalent ($730$\,g/cm$^2$).

\subsection{The cryostat}
\label{sectioncryostat} The cryostat consists of three concentric cylinders (see Figures
\ref{photochamber} and \ref{allcutmu}); it is constructed out of non--magnetic stainless
steel (316L) in order not to distort the magnetic field. All the signal cables, the HV
cables, slow control cables (and additional filling tubes) pass through vertical chimneys
($80$\,mm inner diameter and about $500$\,mm long) welded on top of the LAr and LN$_2$
vessels at the end, which is outside the magnet (see Figure~\ref{allcutmu}). A burst
disk, safety valves and an electro--mechanical  valve to regulate the nitrogen pressure
were mounted on top of the
chimneys.\\
The LAr vessel, containing the chamber, has an inner diameter of $250$\,mm and a total
length of $1371$\,mm yielding a volume of about $65\,\ell$ of argon. The chamber is
positioned in the cryostat such, that it is in the center of the magnet. The vessel is
sealed on both ends with CF250 flanges. The liquid argon is filled and emptied through a
$16$\,mm tube at the bottom of the vessel (see Figure~\ref{Fillingfig}). \\
The next concentric cylinder around the LAr vessel contains the liquid nitrogen bath. The
absolute pressure is regulated to $2.7\!\pm\!0.1$\,bar in order to reach the boiling
temperature of about $87$\,K for argon at $1$\,bar, hence preventing the freezing out of
the LAr. Since the LAr vessel is hermetic (the over-pressure valve opens at about
$1.6$\,bar, well above the operating pressure of about 1 bar), the pressure of the argon
is regulated only by its temperature, i.e., the temperature of the bath. The LN$_2$
cylinder is $1434$\,mm long and has an inner diameter of $350$\,mm with a volume of about
$75\,\ell$ of LN$_2$. The technical grade LN$_2$ is contained in a dewar\footnote{Air
Liquide, type Ranger 180--15,5.} with a total capacity of $180\,\ell$ at a pressure of
$4.5$\,bar in order to fill the bath at $2.7$\,bar with a sufficient flux. During the
data taking the LN$_2$ bath had to be refilled every $12$~hours (see chapter \ref{run}). \\
The flanges of the LAr and the LN$_2$ vessels can be heated with two heating
foils\footnote{MINCO Products, Inc. type HK5171R331L12.} with a power of $40$\,W per foil
to speed up the warming process in case the containers have to be opened. \\
Both, the LN$_2$ and the LAr vessel, have to be vacuum insulated. The last concentrical
cylinder serves as insulation vacuum chamber and has an outer diameter of $549$\,mm and a
length of $1917$\,mm with an evacuated volume of about $220\,\ell$. Also the vertical
chimneys of the LN$_2$ and the LAr vessel are vacuum insulated up to the top. These two
tubes and the whole LN$_2$ vessel are wrapped with about $40$~layers of super--insulation
foils to suppress the irradiated energy from the vacuum chamber (which is at room
temperature). The insulation vacuum was better than $10^{-6}$\,mbar with the cold
cryostat.

\subsection{The magnet}
\label{sectionmagnet} The solenoid magnet used in this experiment is the recycled SINDRUM
I magnet from PSI. The bore holes in the end caps of the yoke were widened in order to
have a $580$\,mm
aperture to permit the insertion of the cryostat. \\
The magnet has a cylindrical shape with an outer diameter of the yoke of $1260$\,mm and a
length of $1280$\,mm. The power supply delivers a maximal DC current of of $850$\,A at a
voltage of $260$\,V; the power consumption of the magnet is about $220$\,kW for the
maximal current. Table~\ref{tabmagnet} summarizes the most important parameters of the
magnet and its cooling system.
 \begin{table}
 \begin{center}
 \begin{tabular}{|p{4cm}|p{5cm}|}
 \hline
                          & \O\,$1260$\,mm \\
 External dimensions & $1280$\,mm length \\ \hline
 Weight & 5.5\,Tons \\ \hline
                          & \O\,$580$\,mm \\
 Internal dimensions & $1120$\,mm length \\ \hline
 Magnetic field & $0.55$\,T (at $850$\,A) \\ \hline
 Homogeneity of the magnetic field & 1.5\,\% inside the chamber \\ \hline
 Max. current & $850$\,A DC \\ \hline
 Power consumption & $220$\,kW at 850 A \\ \hline
 Temperature difference at maximum power & $40$\,$^\circ$C \\ \hline
 \end{tabular}
 \caption{\label{tabmagnet} Parameters of the SINDRUM I magnet.}
 \end{center}
 \end{table}

 \begin{figure}
 \begin{center}
    \includegraphics[width=12cm]{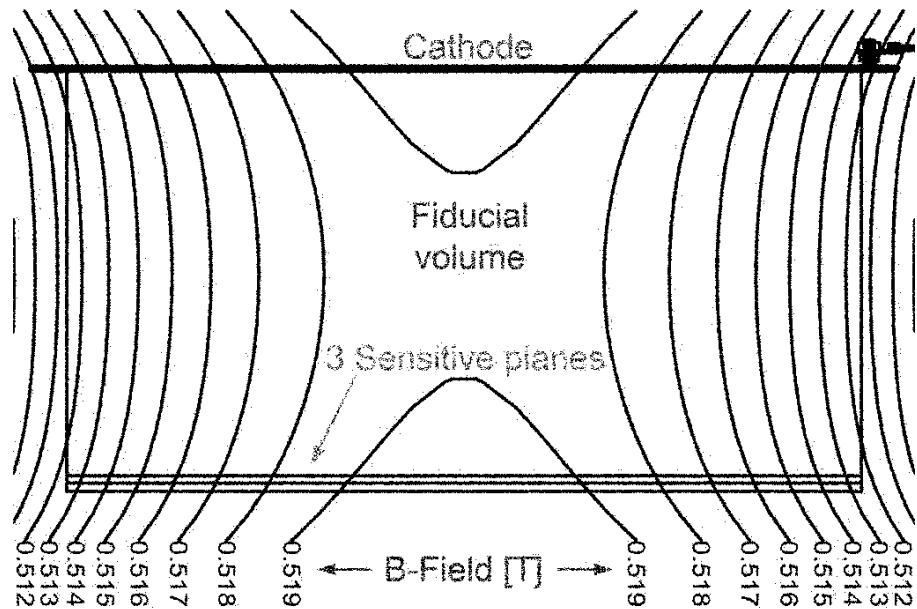}
  \caption[Map of the magnetic field in the time projection chamber.]
 {\label{Mfieldchamber} Map of the magnetic field in the time projection chamber.
 The map is drawn for a horizontal plane through the magnet axis, it has cylinder symmetry.
 The magnet current for the field measurements was $800$\,A ($850$\,A is the maximal current).}
 \end{center}
 \end{figure}
A field map was measured inside the LAr vessel, in the region of the drift chamber; it is
shown in Figure~\ref{Mfieldchamber}. The current for these measurements was $800$\,A; the
magnetic field reaches a maximum of $0.518\,T$ in the center of the volume of the chamber
and about $0.511\,T$ at the side. Extrapolating linearly with the current, this
corresponds to a field of $0.550$\,T in the center with a current of $850$\,A. The
inhomogeneity of the field inside the chamber is of the order of $1.5$\,\%.

\subsection{The trigger system}
\label{triggercounters} 
\begin{figure}
 \begin{center}
    \includegraphics[width=14cm]{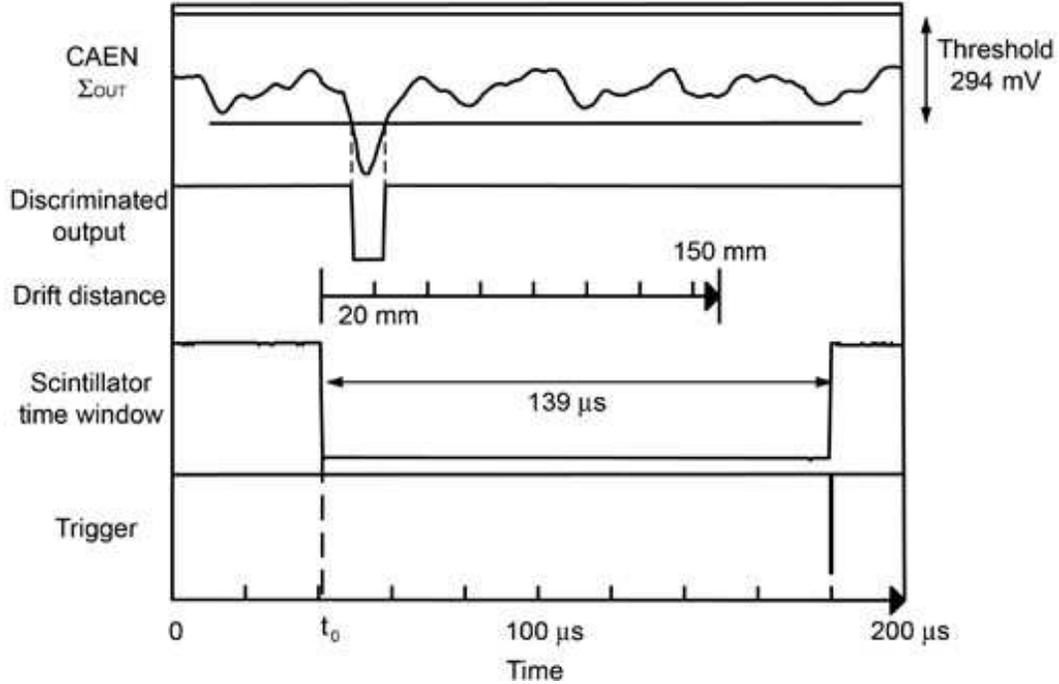}
  \caption[Timing diagram for the signals used to trigger on stopping muons.]
 {\label{triggeroscilloscope} Timing diagram for the signals used to trigger
 on stopping muons. The trigger signal is
 synchronized with the scintillators, in oder to get the $t_o$ needed to determine the drift time.}
 \end{center}
 \end{figure}

There are a total of eleven trigger scintillators mounted as
shown in Figure~\ref{allcutmu}: three scintillators are on the top of the magnet, two are
in the bore hole on top of the cryostat and six are at the bottom of the magnet. The
plastic scintillators measure $822$\,mm$\times$\,$155$\,mm and are $5$\,mm thick; they
are read
out at one end by a Philips photomultiplier\footnote{Philips, type XP2262.} tube. \\
The angular distribution of cosmic muons on the surface of the earth is proportional to
$cos^2 \theta_Z$ ($\theta_Z$ is the zenith angle), hence, most particles come from the
zenith and the flux decreases with increasing zenith angle; this is even enhanced by the
surrounding material of the detector. With a sensitive area of $0.045$\,m$^2$ and a solid
angle acceptance around the zenith direction of $0.28$\,sr, the estimated muon rate is
about $0.7$\,Hz \cite{dissmarco}; this value is consistent with the measured trigger rate
of $0.55$\,Hz. For stopping muons, the estimated rate corresponds to one event every
$58$\,s, which is also compatible with the measured value.\\
The data were taken with two different triggers. The first one tiggers on through--going
muons and requires a coincidence between the scintillators on top, in the bore hole and
at the bottom of the magnet. \\
The second trigger was used to detect stopping muons. The V791 CAEN cards described in
chapter~\ref{readout} have an output of the amplified analog sum of all $32$~channels.
One analog sum of the module processing the signals from the central region of the
collection plane was used for this trigger. The two scintillators in the magnet bore hole
were used to open a gate; if a particle passes through the chamber the analog sum would
be over a given threshold. The coincidence of the two signals generates a trigger at the
end of the gate. The $t_0$ (passage of the particle through the chamber), used to
determine the drift time, corresponds to the leading edge of the gate.
Figure~\ref{triggeroscilloscope} shows the timing diagram for this trigger.

 \subsection{The slow control system}
 \label{secslowcontrol}
The slow control sensors are crucial to control the experiment during the critical
phases, like the pumping period, the cooling down with LN$_2$ and filling LAr, and during
the warming up; it is also important to monitor permanently the experiment during the
data taking phase. The slow control sensors measure the temperature in many positions of
the experiment, inside the liquid argon, around the cryostat and also in the cooling
system of the magnet. Pressure gauges measure the argon and the nitrogen pressure and the
capacitive level meters measure the level of the LN$_2$ and the LAr. All the data are
read by a dedicated computer in a cycle of about $30$\,s and stored in a file (see
\cite{dissmarco} for more details).

\subsection{Purification of the LA\MakeLowercase{r} during the filling}
\label{secpurification}
 \begin{figure}
 \begin{center}
   \includegraphics[width=12.5cm]{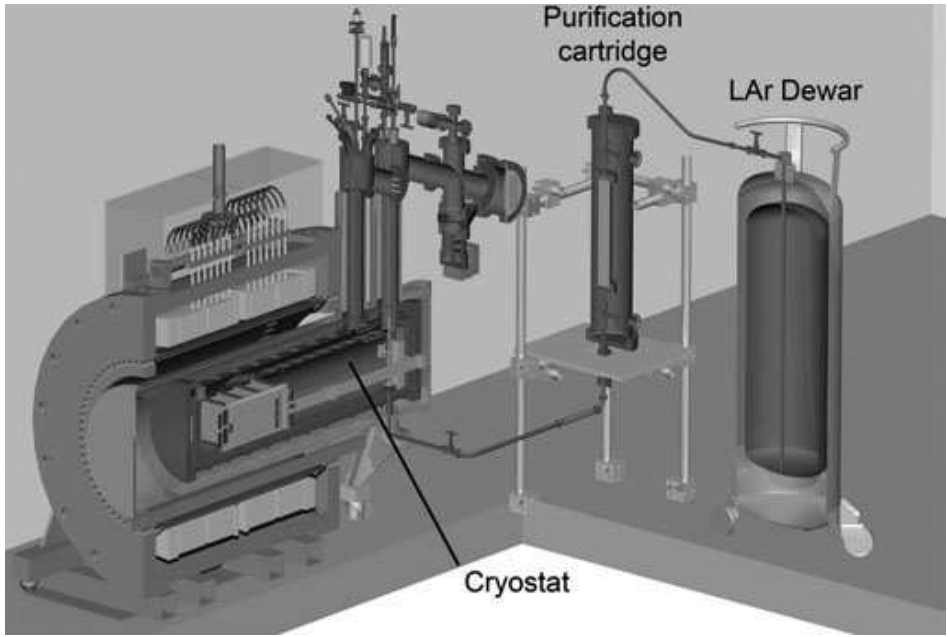}
 \caption[The setup with the purification cartridge.]
 {\label{Fillingfig} The setup with the LAr dewar and the purification
 cartridge used to fill the LAr vessel with the chamber.}
 \end{center}
 \end{figure}
To reach an electron live time in liquid argon of $150$\,$\mu$s (corresponding to the
maximal drift time in the chamber), the vessel and the chamber had to be cleaned and
evacuated to avoid electronegative impurities on the surfaces. Pumping the LAr vessel
below $1\cdot10^{-4}$\,mbar guarantees a purity better than $1$\,ppb of residual
oxygen. In our case, the chamber and the cables were pumped to $5\cdot10^{-6}$\,mbar.\\
The commercial liquid argon 46\footnote{Purchased from Carbagas AG} has a declared
content of $\mathrm{O_{2}}\!<\!5$ ppm and $\mathrm{H_{2}O}\!<\!10$ ppm and there is no
commercially available argon with a purity of the order of ppb. Thus, a filter for
electronegative ions has been placed between the dewar with the commercial LAr and the
vessel with the
chamber as shown in Figure~\ref{Fillingfig}.\\
The filter consists of a cylinder with $63$\,mm internal diameter and $463$\,mm length
made of stainless steel, filled with fine (grain size $\approx\!5$\,$\mu$m) copper
powder. On both ends of the tube there are metal sieves with $\approx\!3$\,$\mu$m holes
in order to contain the powder inside the cartridge. The small amount of oxygen content
in the liquid argon will be bound chemically by the copper in the cartridge via the
$\mathrm{2Cu+O_2\rightarrow 2CuO}$ reaction. At $87$\,K the water inside the LAr is
trapped on the surface of the powder and is easily filtered.\\
The cartridge is enclosed by an insulated cylinder of $200$\,mm inner diameter and
$898$\,mm length. Before filling with LAr, the cartridge is cooled by inserting LN$_2$
into this cylinder.\\
The cartridge was first filled with copper oxide ($\mathrm{CuO}$) purchased from
Fluka\footnote{Fluka, Cupric oxide, puriss, p.a. $\geq\!99$\%} and then reduced by
flowing a hydrogen/argon gas mixture at about $200$\,$^\circ$C through the closed
cartridge; the chemical reaction of the reduction process is $\mathrm{CuO +
H_2\rightarrow Cu + H_2O}$. The admixture of argon gas avoids an overheating due to the
reaction and the temperature can be controlled by varying the argon percentage, it also
helps the water vapor to be expelled.

\section{Test run with cosmic rays}
\label{run} In November 2004 the setup was ready for a first test. Starting with the
vacuum pumping of the liquid argon vessel, the slow control program stored all important
information on the status of the setup in a completely automatic way.

\subsection{Vacuum in the cryostat}
\label{vacuum} 
\begin{figure}
 \begin{center}
   \includegraphics[width=14cm]{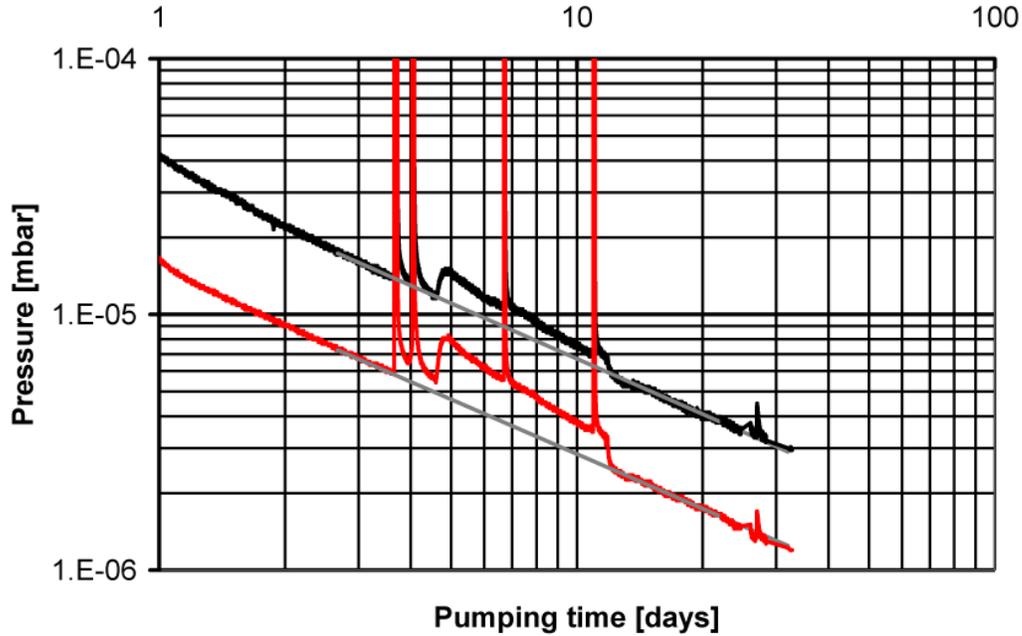}

 \caption[Pressure measurements in the LAr cryostat during the pumping period.]
 {\label{vacuumfig} Pressure measurements with two vacuum gauges on the LAr cryostat
 during the pumping period of a month. The
 pressure near the drift chamber in the cryostat should be between the two measurements.
 The zero point of the time coordinate corresponds to the 16$^{th}$ of September 2005 (see
 text).}
 \end{center}
 \end{figure}

The liquid argon vessel with the chamber installed was pumped the first
time in May $2004$ to test the vacuum tightness of the vessel and to let the cables (with
a total length of about $3$\,km) out-gas. In the next month the vessel was kept as much
as possible under vacuum or under argon or dry nitrogen gas to avoid that the cables and
the cryostat surfaces absorbed water. The pressure reached shortly before the liquid
argon filling was better than $5\!\times\!10^{-6}$\,mbar, well below the requirement. In
Figure~\ref{vacuumfig} the pressure during the period from the 16$^{th}$ of September
until the 19$^{th}$ of October 2004 is shown. During this pumping period, two vacuum
gauges mounted on the LAr vessel were read: one near the turbomolecular pump, the other
one was mounted on the filling tube, at the bottom of the vessel; the pressure near the
drift chamber should be between the two values. The four spikes seen in the figure
correspond to vacuum breaks with Ar gas in order to finish the mounting. The vacuum
breaks had no influence on the final vacuum, and after a few hours the equilibrium with
the out-gassing was reached again. Around day $5$ until day $11$ the temperature of the
vessel was slightly increased up to about $35$\,$^\circ$C, as a consequence, the
out-gassing rate and the pressure increased. After this period the temperature returned
to room temperature of about $21$\,$^\circ$C.

The pressure reached $1\!\times\!10^{-6}$\,mbar with the cryostat at room temperature and
$1\!\times\!10^{-7}$\,mbar with the liquid nitrogen vessel filled, due to the big cold
surface acting as a cryogenic pump for water.

\subsection{Cooling down and filling with liquid argon}
\label{filling} 
\begin{figure}
 \begin{center}
   \includegraphics[width=14cm]{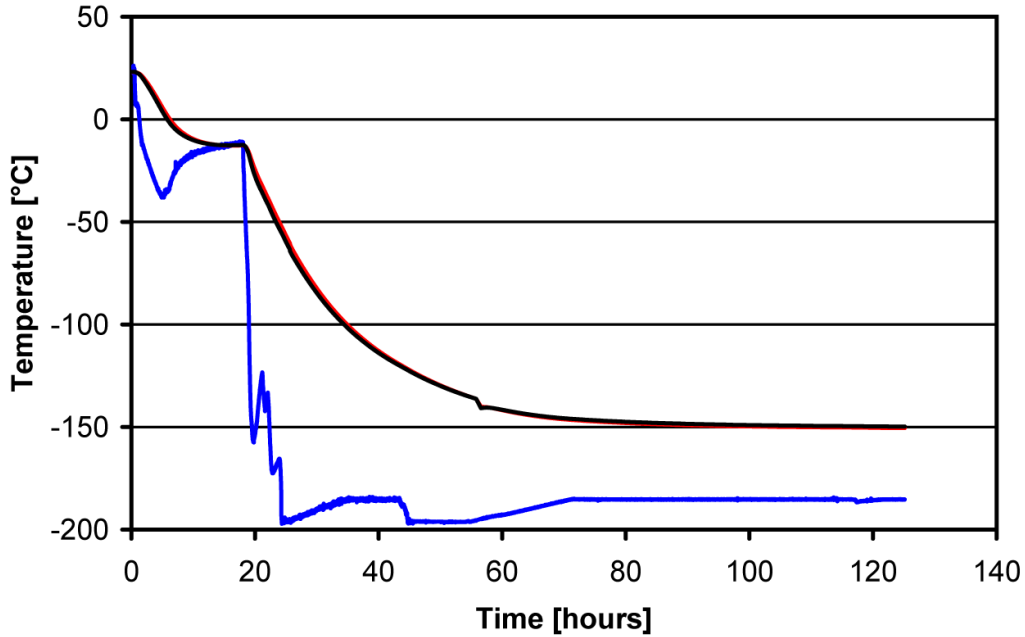}

 \caption[Temperature of the drift chamber and the LAr vessel during the cooling down period]
 {\label{coolingdownfig} Temperature at the bottom of the LAr cylinder (blue),
 on the drift chamber frame (red) and the ambient
 temperature in the LAr vessel (black) during the cooling down period.}
 \end{center}
 \end{figure}

\begin{figure}
 \begin{center}
   \includegraphics[width=12cm]{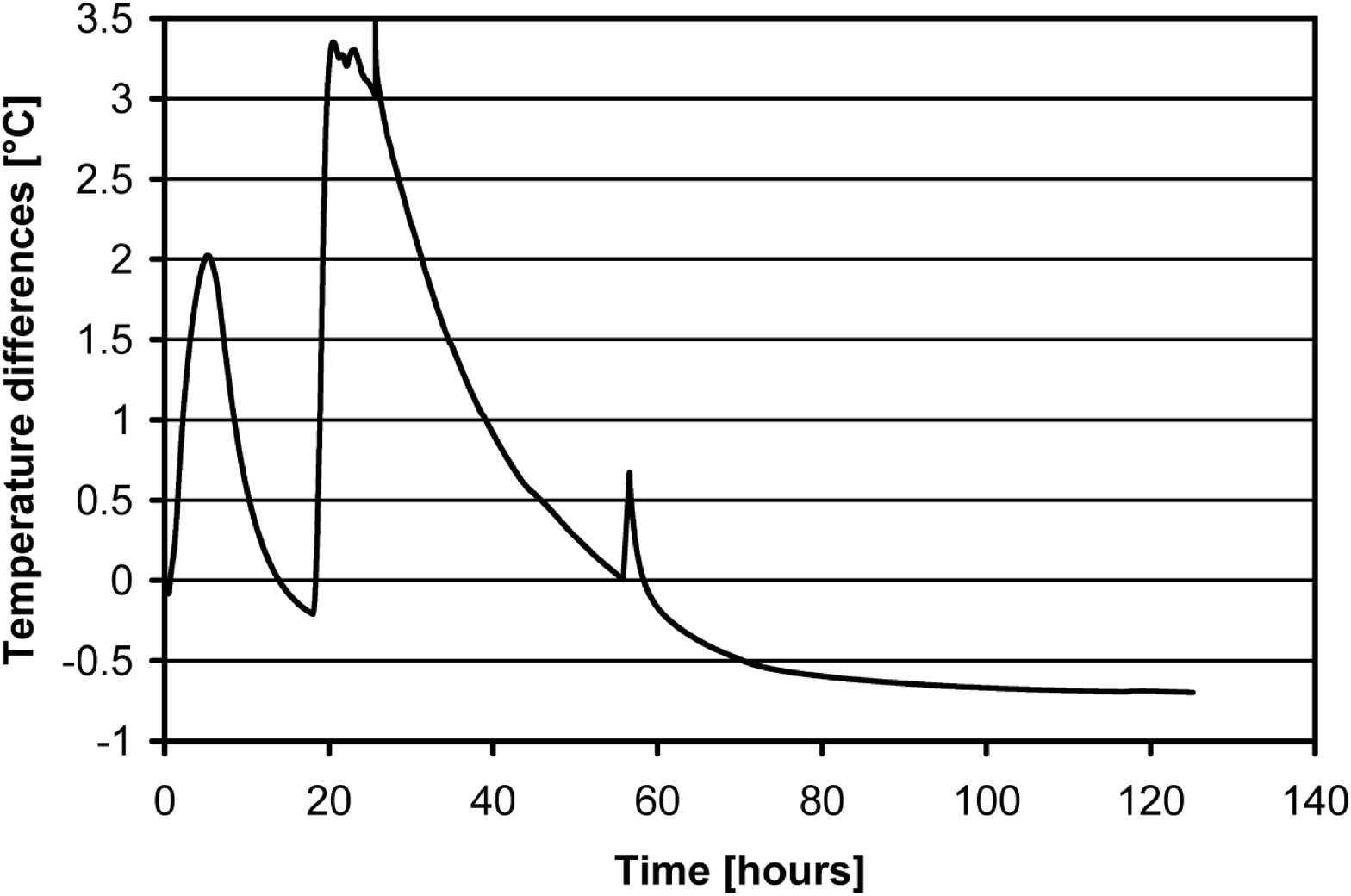}
  \caption[Temperature difference in the chamber during the cooling down.]
 {\label{coolingdowndif} Difference between the temperature of the drift chamber frame and the ambient
 temperature in the evacuated LAr vessel during the cooling down period.}
 \end{center}
 \end{figure}

In November 2004 the liquid nitrogen vessel was cooled down slowly by
filling small amounts of liquid nitrogen distributed over a period of about two days; the
LAr vessel was still under vacuum during this phase. The filling rate was chosen in order
to keep the temperature of the drift chamber as homogeneous as possible.
Figure~\ref{coolingdownfig} shows the temperature at three different locations: at the
bottom of the LAr cylinder (blue), on the drift chamber frame (red) and the ambient
temperature in the LAr vessel (black); the latter two temperatures are almost
indistinguishable on Figure~\ref{coolingdownfig}. The difference between the temperature
of the drift chamber frame and the ambient temperature in the LAr vessel is shown in
Figure~\ref{coolingdowndif}; the liquid nitrogen filling rate was chosen in order not to
exceed a maximal difference of $4$\,$^\circ$C.

Once the LN$_2$ vessel was filled with liquid nitrogen, the pressure was regulated (see
chapter~\ref{setup}) to an absolute pressure of $2.7$\,bar (in order to reach the right
temperature for liquid argon with an absolute pressure of about $1$\,bar), and the LAr
container was filled ($\sim\!65$\,$\ell$) in about $2$~hours.
 \begin{figure}
 \begin{center}
 \setlength{\unitlength}{1mm}
 \begin{picture}(125,200)
    \put(1,135){\includegraphics[width=100\unitlength]{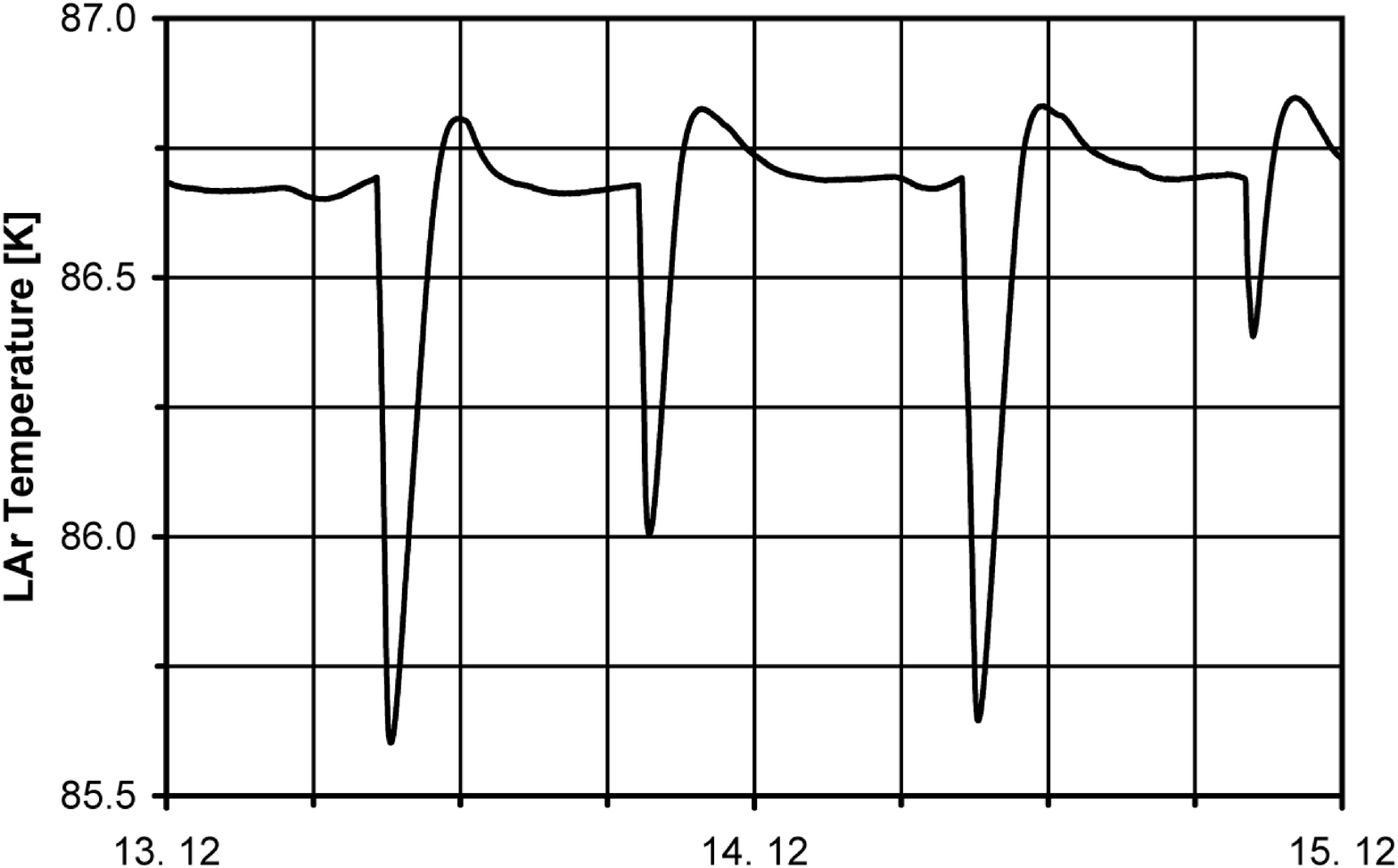}}
   \put(2,67){\includegraphics[width=98\unitlength]{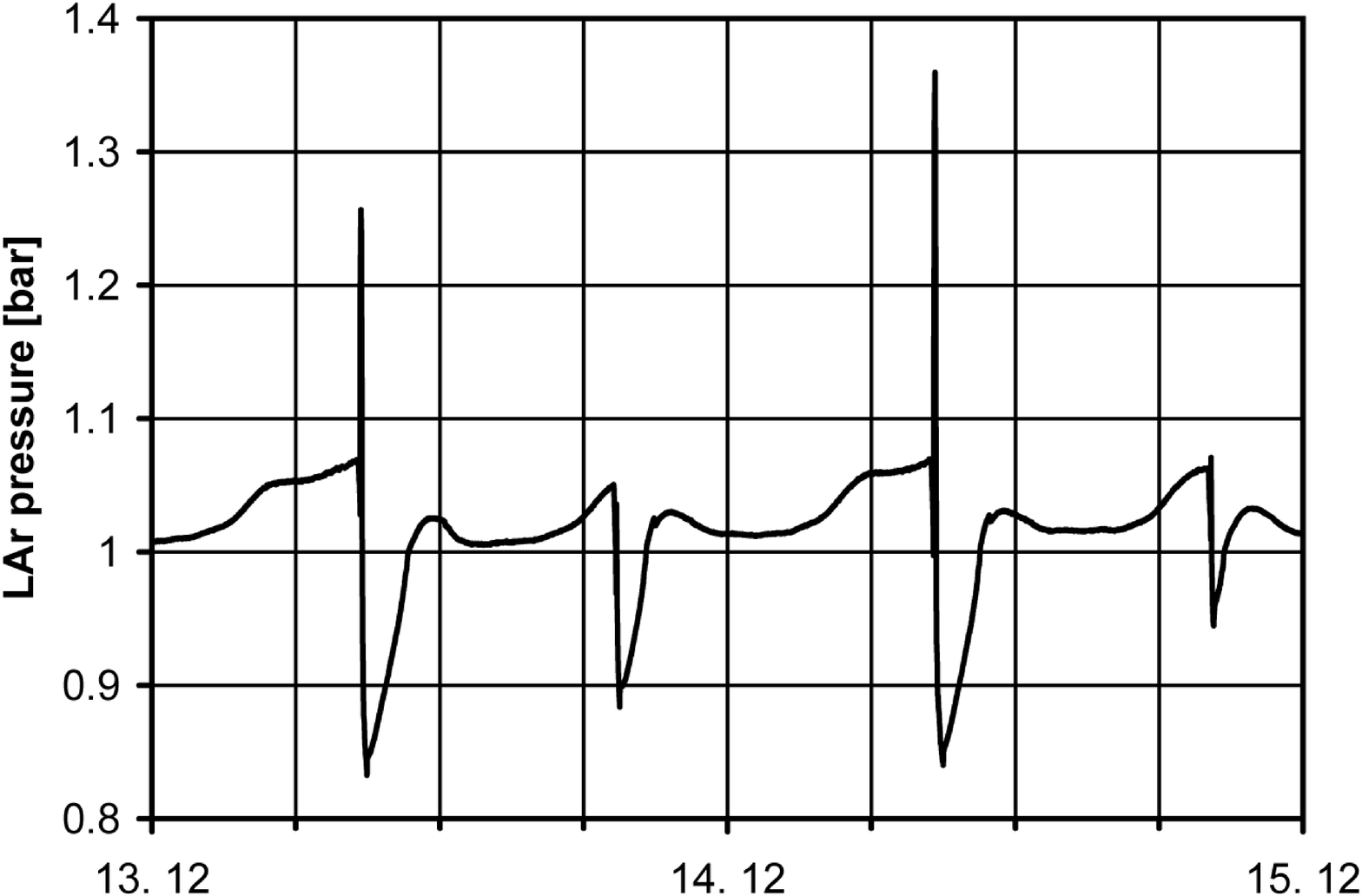}}
   \put(0,0){\includegraphics[width=100\unitlength]{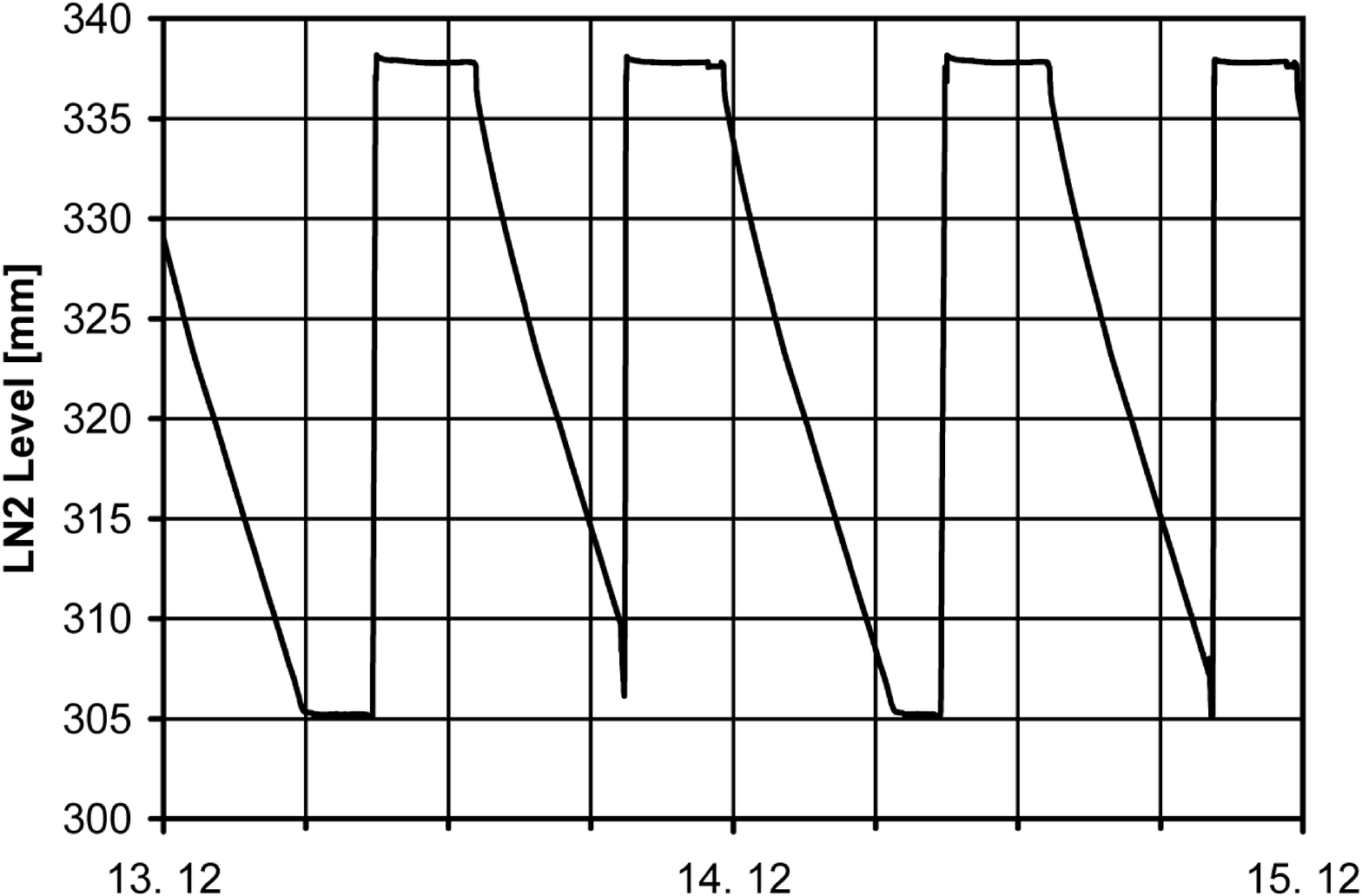}}
  \end{picture}
 \caption[Temperature and pressure of the LAr during two days of data taking.]
 {\label{measureLAr} LN$_2$ level and the temperature and pressure of the liquid
 argon during two days of data taking. Visible are the liquid nitrogen fillings twice a day.}
 \end{center}
 \end{figure}

\subsection{Measurements with the LA\MakeLowercase{r} TPC}
On November 23$^{rd}$ 2004 the setup was ready for a first test. Starting the tests
without magnetic field and triggering with the scintillator counters, clean cosmic ray
tracks were immediately observed at a drift field of $500$\,V/cm. With every run
$100$~trigger events were stored. At the beginning there was a commissioning period
without magnetic field, when some through--going muon events were taken with increasing
drift field. The maximum field applied was $1.5$\,kV/cm, i.e., $22.5$\,kV at the cathode.
After a week of operation, breakdowns occurred in the high voltage system. After this
time the high voltage was limited to $4.7$\,kV, corresponding
to a drift field of $0.3$\,kV/cm. \\
The level of the liquid nitrogen was monitored to avoid a drop to a too low level, which
would cause an increase of the temperature in the liquid argon vessel. A re-filling every
$12$ hours was necessary to replace the evaporated nitrogen, corresponding to a global
consumption of about $1.4$\,$\ell$/h, equivalent to $62$\,W of thermal losses.
Figure~\ref{measureLAr} shows the LN$_{2}$ level and the temperature and pressure of the
liquid argon during two days of data taking with the visible liquid nitrogen fillings
twice a day.

In the beginning of the test run a coincidence of the scintillators was required for a
trigger to detect through--going muons. In this case only about $20$\,--\,$30$\,\% of the
triggered events had a track in the drift chamber. This efficiency is given by the
position of the scintillators, which give a trigger signal also for muons passing close
to the chamber, but not traversing the sensitive volume. To trigger on stopping muons, a
coincidence between the scintillators on top of the cryostat in the magnet bore hole and
the analog sum output of the 32 channels from a CAEN V791 board was required. The
scintillators give in this case only the $t_0$ time of the event, needed to determine the
drift time.

\begin{figure}
 \begin{center}
 \setlength{\unitlength}{1mm}
 \begin{picture}(130,198)
   \put(3,3){\includegraphics[width=60\unitlength]{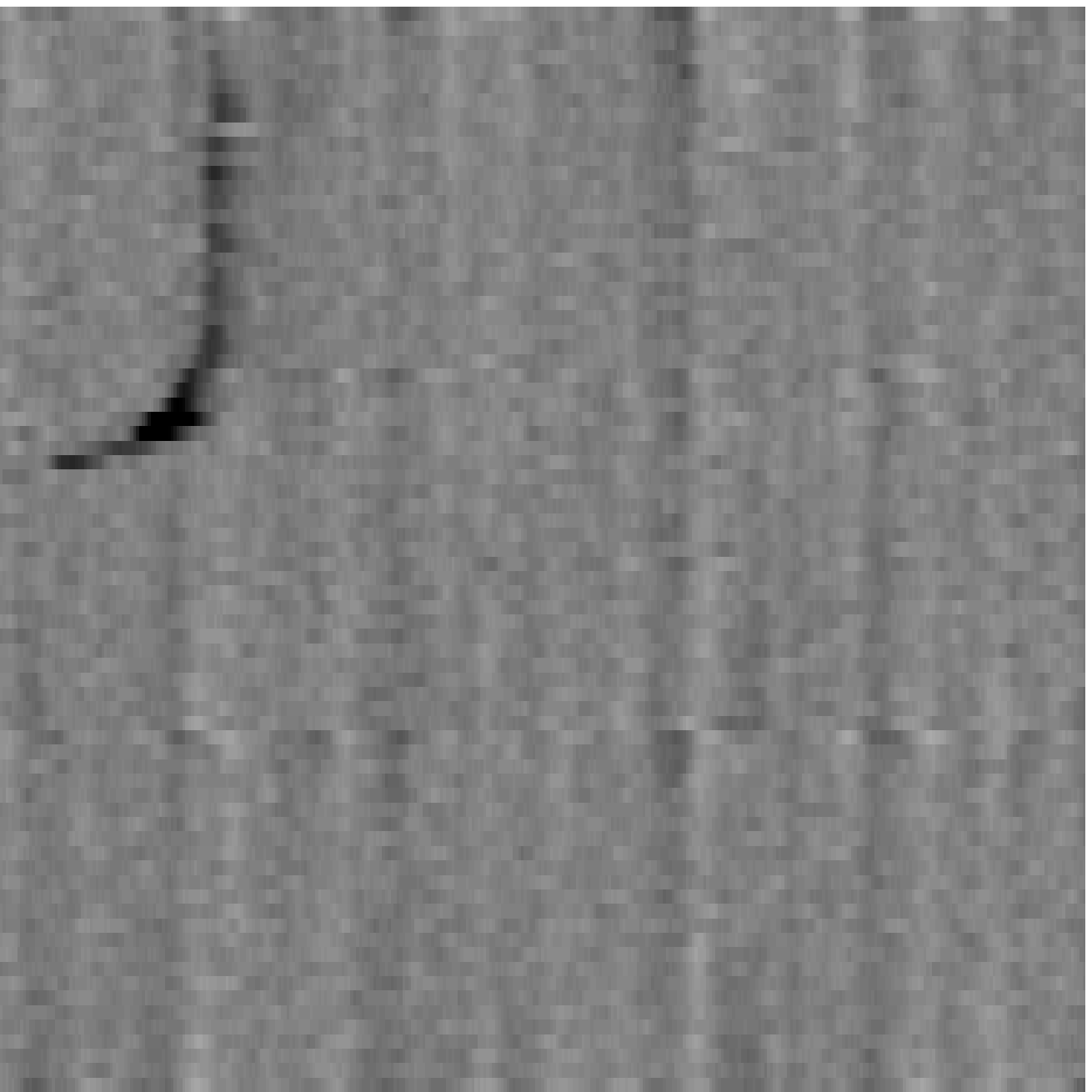}}
   \put(65,3){\includegraphics[width=60\unitlength]{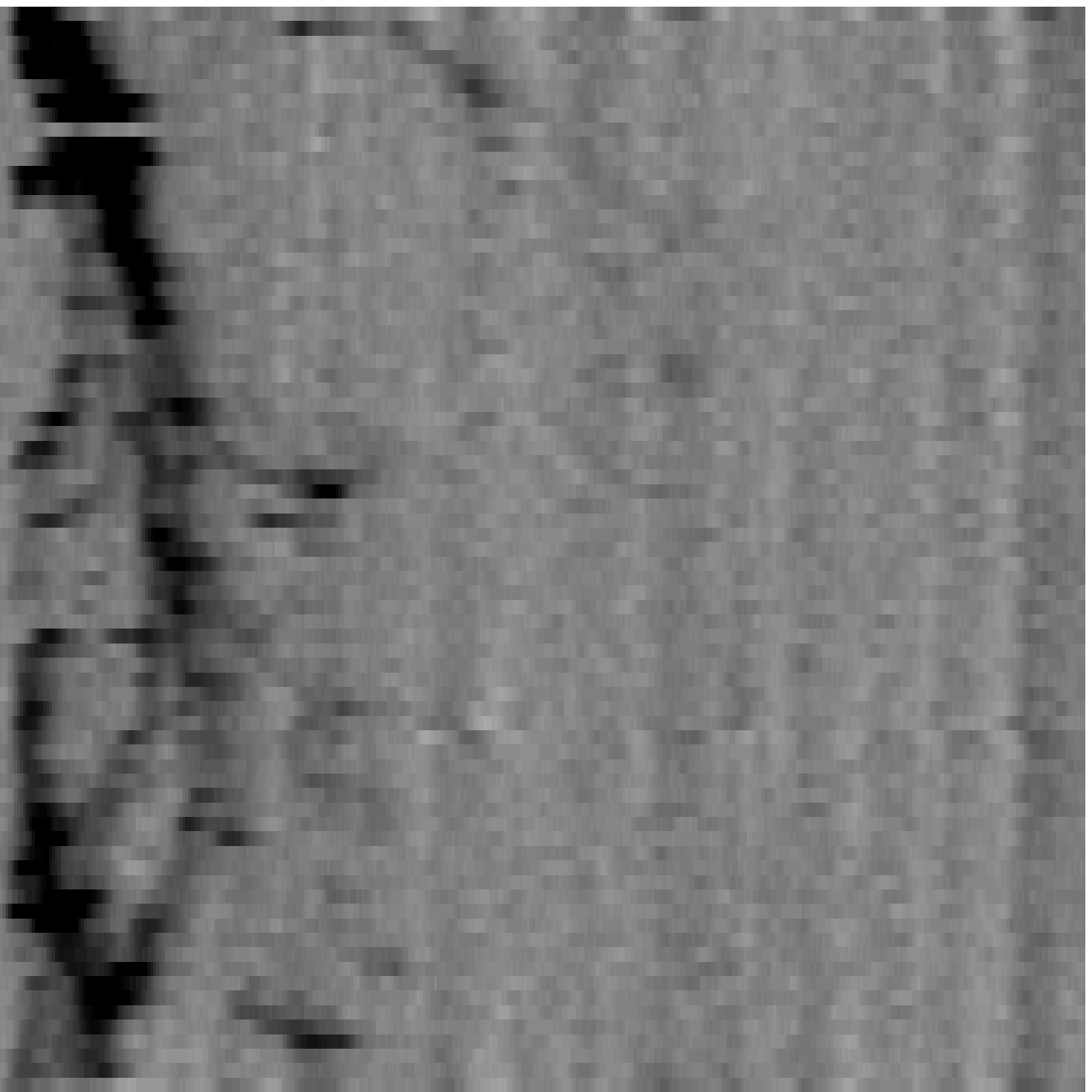}}
   \put(3,65){\includegraphics[width=60\unitlength]{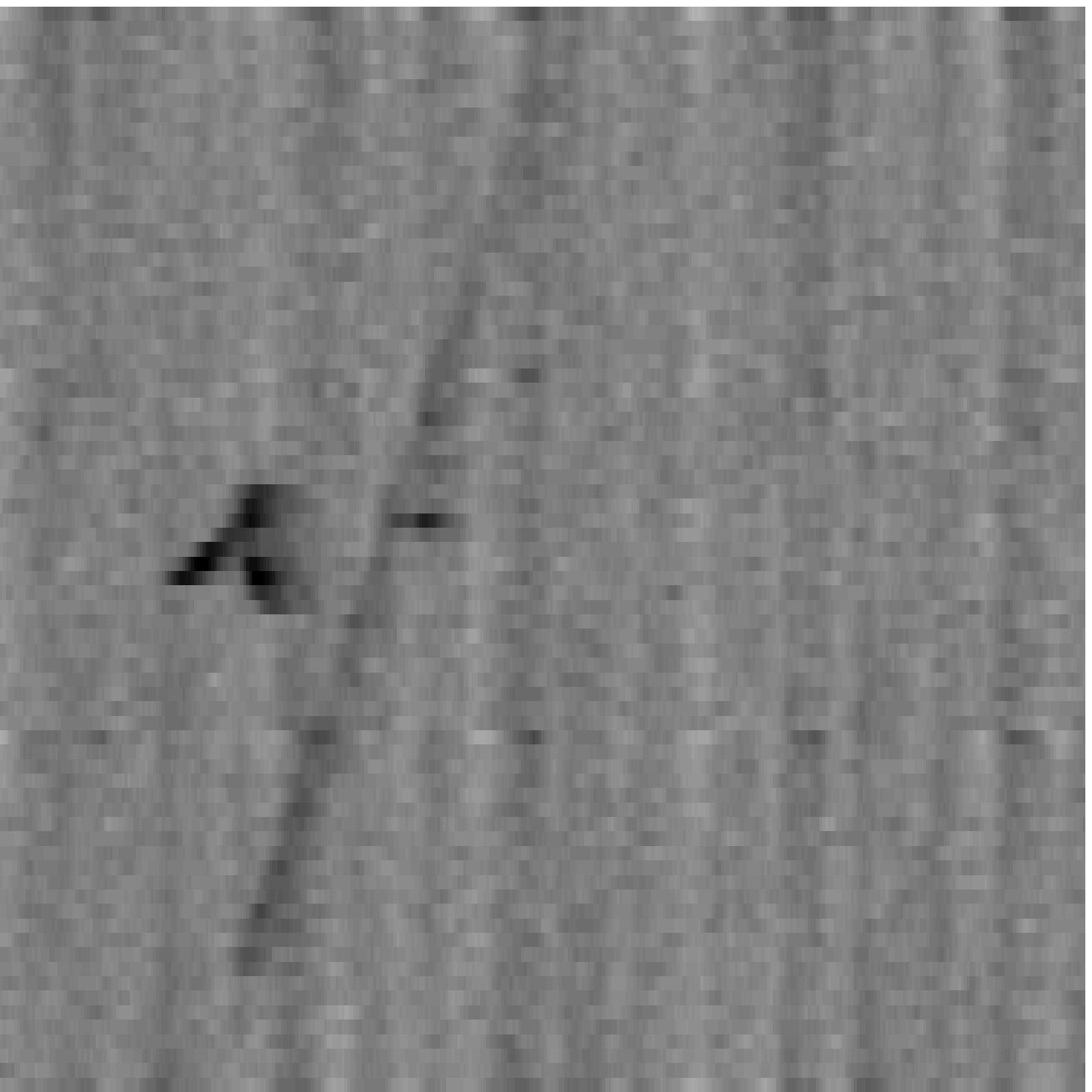}}
   \put(65,65){\includegraphics[width=60\unitlength]{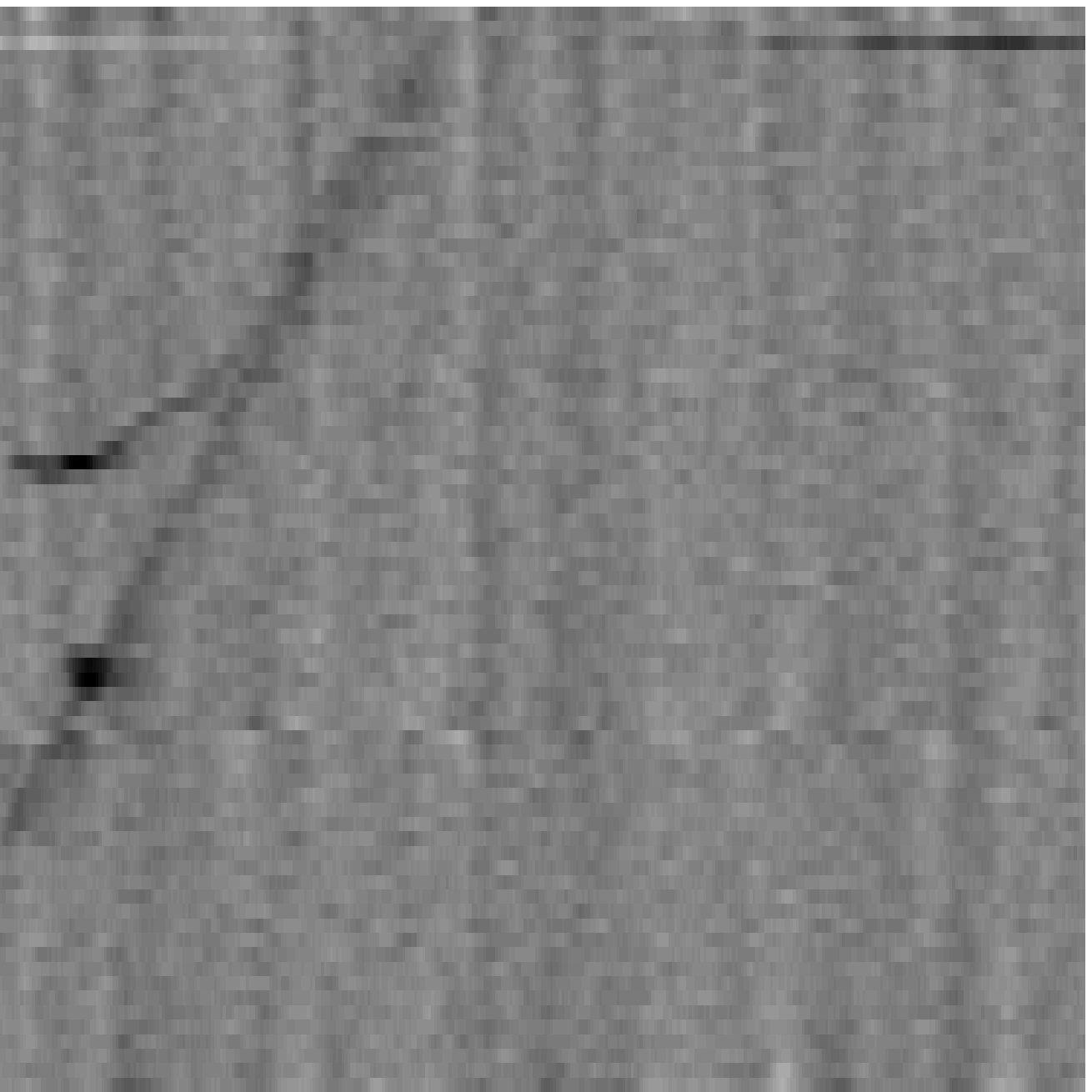}}
   \put(3,127){\includegraphics[width=60\unitlength]{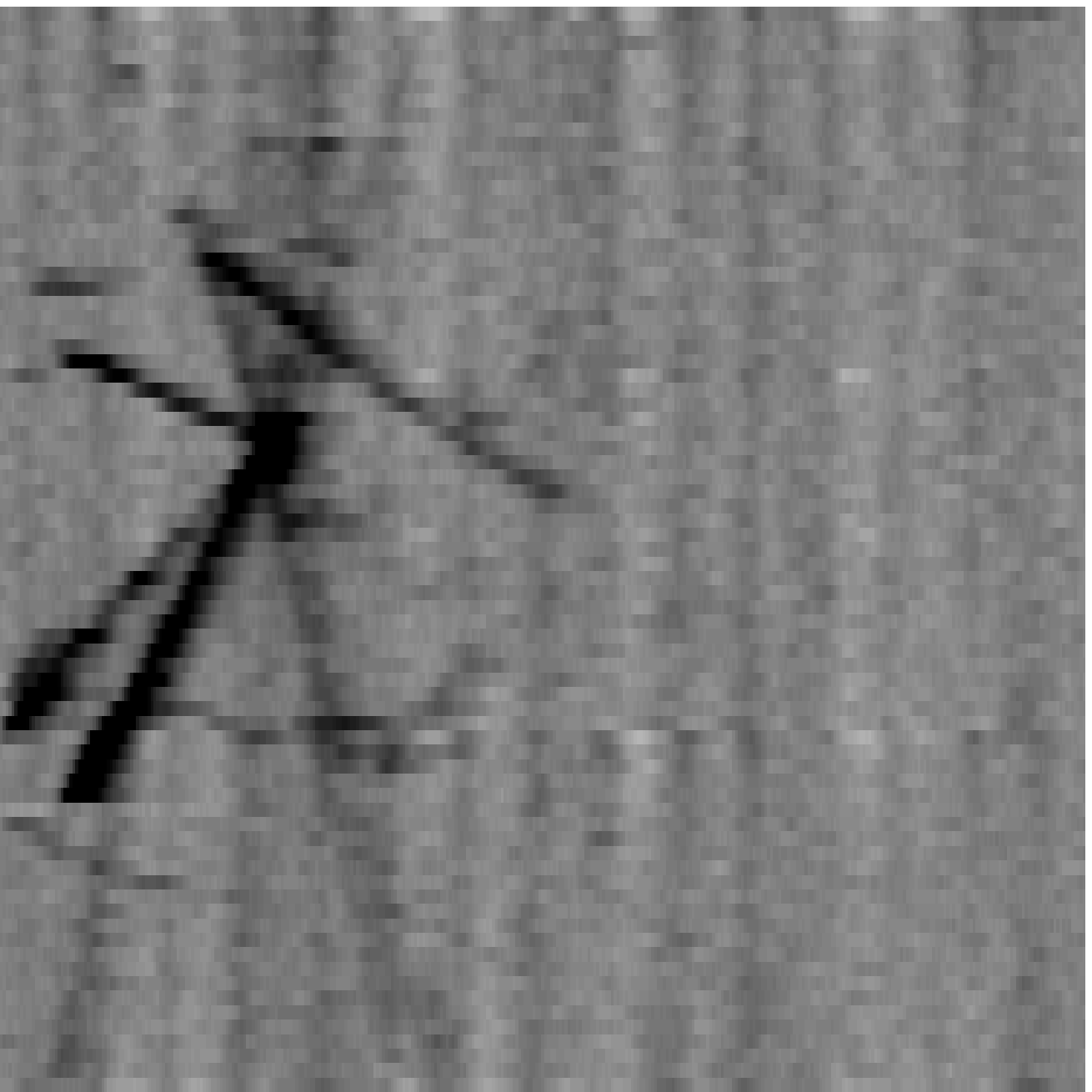}}
   \put(65,127){\includegraphics[width=60\unitlength]{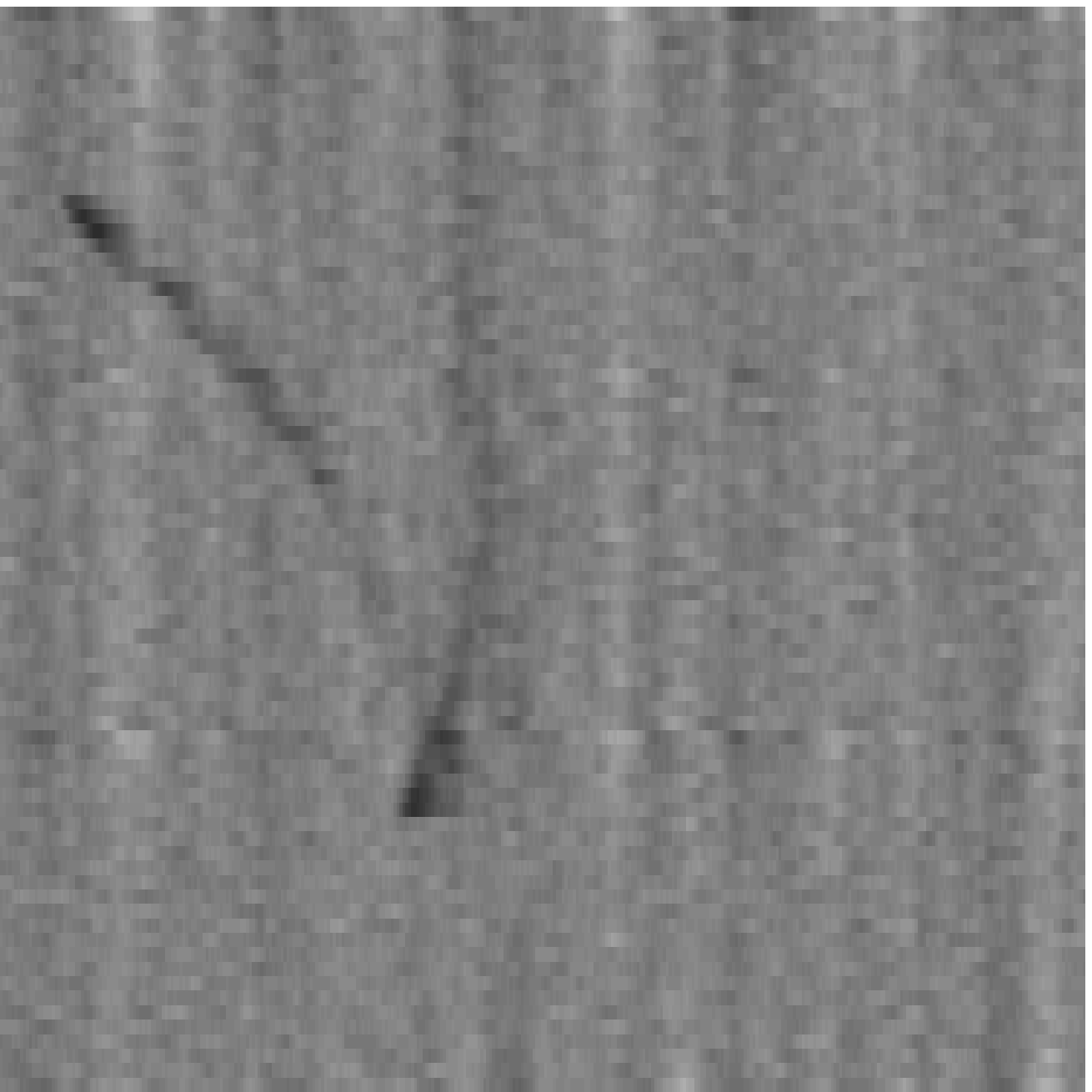}}
   \thicklines
   \put(0,0){\vector(1,0){63}}
   \put(64,2){\makebox(0,0)[tl]{$t_{drift}$}}
   \put(0,0){\vector(0,1){63}}
   \put(-4,14){\rotatebox{90}{Wire number}}

   \put(58,6){\makebox(0,0)[bl]{{\huge e}}}
   \put(120,6){\makebox(0,0)[bl]{{\huge f}}}
   \put(58,68){\makebox(0,0)[bl]{{\huge c}}}
   \put(120,68){\makebox(0,0)[bl]{{\huge d}}}
   \put(58,130){\makebox(0,0)[bl]{{\huge a}}}
   \put(120,130){\makebox(0,0)[bl]{{\huge b}}}

   \thicklines
   \drawline(65,190)(125,190)
   \drawline(65,191)(65,189)
   \drawline(125,191)(125,189)
   \put(95,195){\makebox(0,0)[t]{$\sim\!150$\,mm}}
   \drawline(128,187)(128,127)
   \drawline(127,187)(129,187)
   \drawline(127,127)(129,127)
   \put(130,147){\rotatebox{90}{$\sim\!150$\,mm}}

   \put(33,191){\circle{4}}
   \drawline(32,190)(34,192)
   \drawline(32,192)(34,190)
   \put(33,198){\makebox(0,0)[t]{$\overrightarrow{B}=0.55$\,T}}
 \end{picture}
 \caption[Examples of real events collected.]
 {\label{runevents2} Six examples of real events collected with the liquid Argon TPC prototype
 immersed in a magnetic field of $0.55$\,T. The horizontal axes correspond to the time
 coordinate and the vertical axes are the wire coordinate, both correspond to a full scale of $150$\,mm.}
 \end{center}
 \end{figure}

Figures~\ref{runevents2} shows some examples of visually selected events from the total
sample of $\sim\!30\,400$ collected events. The raw data from the collection plane are
displayed as two-dimensional projection of tracks in the plane perpendicular to the
magnetic field: the horizontal axes correspond to the drift time (converted to a drift
distance with the drift velocity) with a full scale of about $150$\,mm, and the vertical
axes are the strip number coordinates, also with a full scale of $150$\,mm. The magnetic
field is pointing into the plane of the figure and had a value of $0.55$\,T. The events
are interpreted as a) a reaction with an argon nucleus, b) a stopping positive muon with
the decay positron, c) a muon track with an electron--positron pair, d) a muon with a
$\delta$--electron, e) a stopping negative muon and f) an electro-magnetic shower.

\section{Event selection and results}
\label{results} For a first analysis events with a simple topology, namely positrons
emitted by stopped muons and $\delta$-electrons from passing--through muons were chosen
\cite{amuller05}. The collected events were scanned by eye with the Qscan program. This
program displays a two dimensional gray scale representation of the output signal in the
wire/drift time coordinate plane and provides the possibility to call a hit finding and
fitting algorithm, which gives the needed information for the event reconstruction.
Events were selected if the track length of the electron/positron (in the plane
perpendicular to the magnetic field) is at least 2~cm and if the curvature of the track
is well visible, i.e. the transverse projected track should be recognizable as a section
of a circle,
which biases the event selection. \\
A 3D reconstruction of the electron/positron tracks is necessary for a full calorimetric
reconstruction and a determination of the total momentum. A detailed description of the
spatial reconstruction procedure is given in ~\cite{icarus04,Rico02}.

\begin{figure}
\begin{center}
\includegraphics[width=1\textwidth]{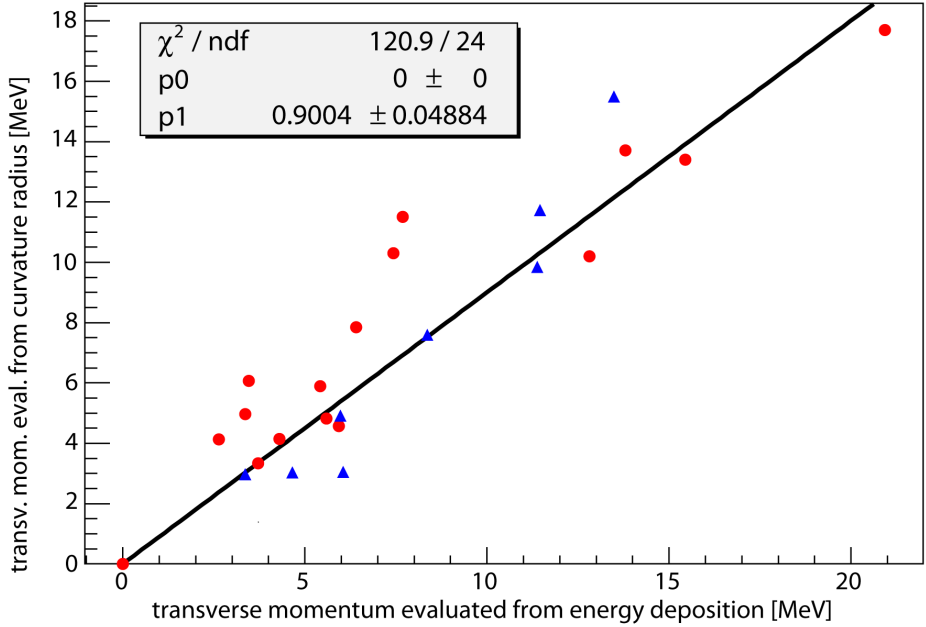}
\end{center}
\caption{Correlation between the two momentum measurement methods for $\delta$-electrons
(circles) and decay-positrons (triangles)} \label{fig:both}
\end{figure}

\begin{figure}
\begin{center}
\includegraphics[width=1\textwidth]{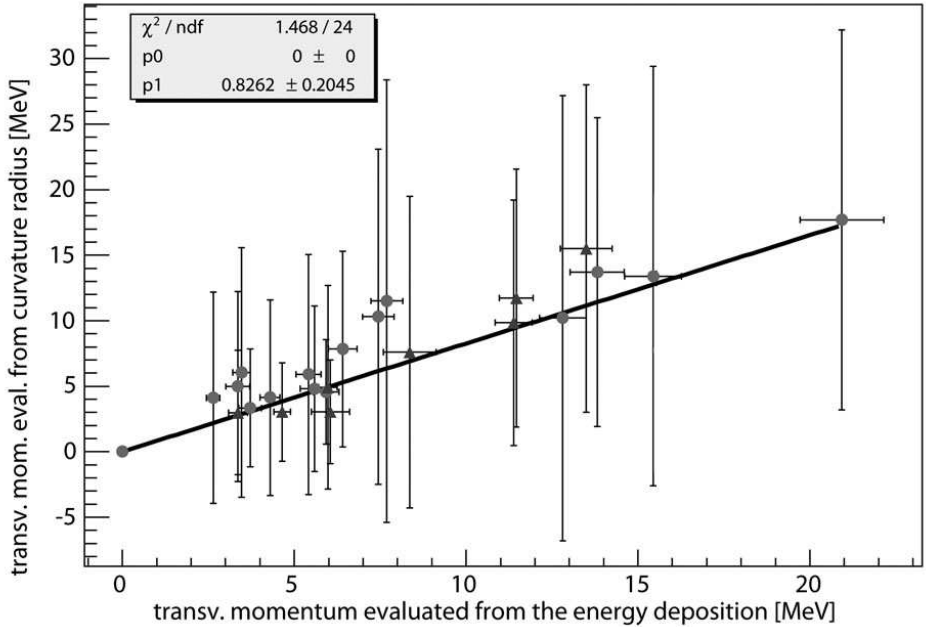}
\end{center}
\caption{Correlation between the two momentum measurement methods for $\delta$-electrons
(circles) and decay-positrons (triangles) with an error estimation due to multiple
scattering} \label{fig:both-error}
\end{figure}

A small sample of 15 $\delta$-electrons and 9 decay positrons, fully contained in the
chamber, were selected from 30'400 recorded and scanned events. These events were
reconstructed and their momentum and kinetic energy was determined from the magnetic
bending and the summed up energy loss, respectively \cite{amuller05}. \\
In Figures~ \ref{fig:both} and \ref{fig:both-error} the transverse momentum obtained from
the energy (charge) measurement is plotted against the momentum from the magnetic bending
for $\delta$-electrons and decay positrons together (circles: $\delta$-electrons,
triangles: decay-positrons). The straight line fit (fit function: $y=p1\cdot x$) through
the data points is also shown. In Figure~\ref{fig:both} no errors were assigned to the
individual data points and Figure~\ref{fig:both-error} includes the errors of the
transverse momentum due to multiple scattering estimated according to equation
\ref{momentumresolution}. These errors are obviously much larger than the scattering of
the data points around the fitted straight line; the $\chi^{2}/ndf$ is equal 0.05,
indicating that the error is over estimated. This discrepancy can be attributed to the
event selection, introducing a bias towards tracks with little multiple scattering
showing nicely bent tracks.  For the error of the transverse momentum evaluated from the
energy see \cite{amuller05}. \\
The slopes of the two plots are p1=0.9480$ \pm$ 0.0431 for Figure~\ref{fig:both}, and
p1=0.9122 $\pm$ 0.1814 for Figure~\ref{fig:both-error}; thus, the slopes are compatible
with 1.

\subsection{Energy loss and the electron lifetime in LAr}
\label{sec:dEdx} 
\begin{figure}
\begin{center}
\includegraphics[width=1\textwidth]{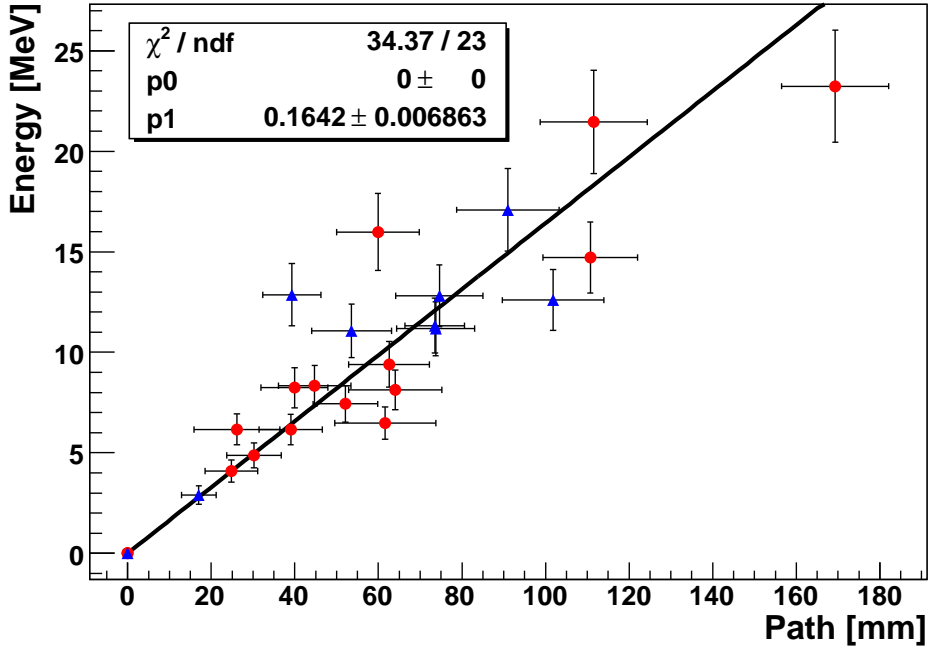}
\end{center}
\caption{Measured energy as a function of the reconstructed path length of the
$\delta$-electrons (circles) and the decay-positrons (triangles)} \label{fig:energypath}
\end{figure}

Figure \ref{fig:energypath} shows the measured kinetic energy plotted
versus the path length. The slope of a linear fit gives an estimate for the (average)
specific energy loss dE/dx, since the energy loss is assumed to be constant (except for
the last 2~mm of a track \cite{amuller05}). The slope in Figure~\ref{fig:energypath}
corresponds to an average energy loss of 1.642 ($\pm$ 0.069) MeV/cm. The theoretically
expected value is about 1.9 - 2.2 MeV/cm for electrons (positrons) in the range between 1
MeV and 20 MeV. Thus, the measured and the expected values differ by about 20\%. However,
the energy loss dE/dx from the measured charge depends on the assumed lifetime $\tau$ of
the drift electrons; a first estimate of 150~$\mu$s was used for the lifetime to
calculate the kinetic energy. Varying the electron lifetime and recalculating the kinetic
energy of all events, the comparison of the average dE/dx with the expected theoretical
value yields an electron lifetime in the range of 60~$\mu$s - 80$\mu$s.

\section{Conclusions}
\label{conclusions} In an effort to further develop the LAr TPC technology, a small test
TPC was built and operated for the first time in a magnetic field ($0.55$\,T), which was
perpendicular to the electric drift field. The quality of cosmic ray tracks is not
significantly decreased with the magnetic field turned on.

Combining the excellent imaging and calorimetric properties of a LAr TPC with a magnetic
field opens new experimental possibilities. The magnetic bending of charged particles
allows the momentum determination also for particles leaving the sensitive volume and the
determination of the sign of the electric charge. The latter feature is a must in future
neutrino experiments searching for CP--violating effects in the lepton sector.

A special interface between the ICARUS front-end electronics (V791 CAEN boards) and the
computer was developed and built for the data acquisition system of this experiment. It
is able to store the continuous flow of digital data in a circular buffer and to send the
data to the PCI card in the computer, once a trigger has occurred. The buffer size is
large enough to store the digitized data of all the channels for a time interval
corresponding to the maximal drift time occurring in the chamber.

A small sample of selected $\delta$-ray and muon decay events were analyzed, showing that
the momentum of an electron (positron) determined from the magnetic bending and from its
kinetic energy are consistent for these events. The kinetic energy was calculated from
the measured ionization charge, summed along the whole track.

\ack We thank PSI for lending us the SINDRUM I magnet with the power supply and ETH
Zurich to provide us with the necessary infrastructure to operate the magnet at ETHZ. We
are also indebted to the INFN Padova group who kindly lent us the readout electronics;
especially we thank Sandro Centro (INFN, Padova) for his support. We thank P. Picchi and
F. Pietropaolo for useful discussions. This work was supported by ETH Zurich and the
Swiss National Science Foundation.

\end{document}